\documentstyle[psfig,epic,eepic,pstricks,pstricks,pst-plot,
               epsf,pst-node,12pt,twoside,fleqn,espcrc2]{article}
\newcommand{\AmS}{{\protect\the\textfont2
  A\kern-.1667em\lower.5ex\hbox{M}\kern-.125emS}}

\title{\fontsize{16}{16.2}\selectfont{Density Functional Theory of Epitaxial
    Growth of Metals}} \author{{\fontsize{14}{14.2}\selectfont{P. Ruggerone,
      C. Ratsch, and M. Scheffler}}
  \address{\fontsize{14}{14.2}\selectfont{Fritz-Haber-Institut der
      Max-Planck-Gesellschaft, Faradayweg 4-6, D-14195 Berlin-Dahlem,
      Germany}}}
\begin{document}
\maketitle
\vspace{.7cm}
\fontsize{14}{14.2}\selectfont{
\begin{abstract}
  \setlength{\baselineskip}{16.5pt} This chapter starts with a summary of the
  atomistic processes that occur during epitaxy.  We then introduce density
  functional theory (DFT) and describe its implementation into
  state-of-the-art computations of complex processes in condensed matter
  physics and materials science. In particular we discuss how DFT can be used
  to calculate parameters of microscopic processes such as adsorption and
  surface diffusion, and how they can be used to study the macroscopic time
  and length scales of realistic growth conditions. This meso- and macroscopic
  regime is described by the {\em ab initio} kinetic Monte Carlo approach.  We
  discuss several specific theoretical studies that highlight the importance
  of the different diffusion mechanisms at step edges, the role of
  surfactants, and the influence of surface stress. The presented results are
  for specific materials (namely silver and aluminum), but they are explained
  in simple physical pictures suggesting that they also hold for other
  systems.
\end{abstract}
}

\vspace{0.7cm}
\fontsize{14}{14.2}\selectfont{
\section{INTRODUCTION}
\label{sec:intro}
\setlength{\baselineskip}{16.5pt}
Many basic concepts for surface diffusion and crystal growth have been
developed already more than 40 years ago (see e.g.
\cite{volmer21,langmuir32,BCF}).
In recent years the subject has attracted significant attention
that is largely
due to new experimental advancements, progress in theoretical methods, and the
importance of crystal growth for technological applications.
A thorough knowledge of the motion of atoms 
at surfaces is a key factor to the understanding of chemical reactions at
surfaces, of crystal growth, and to the question under which conditions 
thermal equilibrium can be achieved (at least locally) at crystal surfaces.
In typical growth
experiments
deposition rates are of the order of a monolayer (ML) per minute,
and the diffusion of atoms at the surface is too slow (at least for
some processes) so that thermal equilibrium is often not reached.
As a consequence, the structures which occur at surfaces 
are usually a result of the kinetics.

Nevertheless, under certain conditions
the resulting structures are ruled by thermal equilibrium, i.e.,
they correspond to the minimum of the free energy.
This may occur 
when the adatom mobility is very high,
the deposition rate is low, or when growth is interrupted and the sample
annealed.
Concerning the acquired
surface morphology one then distinguishes the following 
three different growth modes (cf. Fig.~\ref{growth_modes}):
Frank-van der Merwe or layer-by-layer growth,  
Volmer-Weber growth, where three-dimensional islands are
formed and the overlayer does not
completely cover the exposed substrate surface, and 
Stranski-Krastanov growth 
with layer-by-layer growth supplanted by island growth.
Bauer~\cite{Bauer} had discussed the conditions for these growth modes and
pointed out that the actually realized surface morphology (under thermal
equilibrium conditions) is controlled by the competition of the surface energy
of the substrate, the surface energy of the film, and the interface energy of
the film and substrate. 
\begin{figure}[tb]
\unitlength1cm
\begin{center}
   \begin{picture}(10,9.6)
      \includegraphics{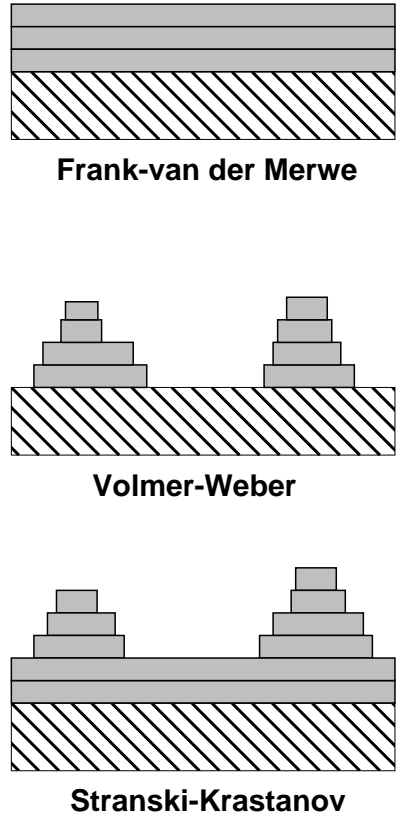}
   \end{picture}
\end{center}
\caption{
Three different growth modes of heteroepitaxial growth.}
\label{growth_modes}
\end{figure}

The modern treatment of growth under non-equilibrium conditions starts with
the seminal paper by Burton, Cabrera, and Frank (BCF)~\cite{BCF}, who realized that
a surface usually is not perfectly flat but has imperfections such as steps.
Steps are the boundary between regions that correspond to upper terraces and
regions that correspond to lower terraces. They might occur at random or can
be created in a controlled manner by cutting the surface in an orientation
close to a low index plane.  In the latter case the surface is called a {\it
  vicinal} surface.  At steps and particularly at kink sites adatoms are bound
most favorable and BCF assumed that on a stepped
surface growth occurs by attachment of deposited adatoms to steps that
subsequently advance. This growth mode is called {\em step flow} and is indeed
easily achieved experimentally at higher temperatures. It results in flat
films.  The step-flow growth mode requires a high mobility of the deposited
atoms, so that they reach the existing steps before meeting other adatoms.
Thus this situation is close to thermal equilibrium.

Usually it is desirable to grow crystals at not too high temperatures. Then
the adatom mobility is lower, and deposited atoms might not reach an intrinsic
step edge. Instead, they will wander around on the surface and meet other
atoms, eventually nucleating a new small island. Further atoms will be caught
at the edges of these islands (or create new ones), so that growth proceeds
via two-dimensional growth of the islands. Additionally, during the initial
stage of film growth the chance that newly deposited atoms land on an island
is small. Nucleation and growth of islands can be described by
phenomenological rate equations \cite{Venables1,Venables2} and we will discuss
this approach briefly in the following Section.  The competition between step
flow growth and growth via nucleation, spread, and eventual coalescence of
islands on the terraces of a vicinal surface can be captured if one
incorporates the rate equation formalism into the BCF equations
\cite{Myers-Beaghton}.

Using rate equations one usually describes diffusion
by {\it effective} parameters but unfortunately
lacks a detailed understanding
of the microscopic mechanism behind them. For example, it has been discussed
by several authors~\cite{bas78,wri78,tun80,kel90,che90,exchange,yu97}
that surface diffusion can occur via two different mechanisms:
\begin{itemize}
   \item [$i)$] An adatom may simply hop from one low-energy site to another
 one, while the substrate reacts only modestly by local relaxations,
or
   \item [ $ii)$] an adatom may diffuse by atomic exchange where 
it changes place with a substrate atom and the ejected substrate
atom moves further.
\end{itemize} 
These two mechanisms are operative at the flat regions of the
surface, but also for diffusion across steps or parallel to step edges.  The
interplay of those different diffusion processes significantly affects the
shape of growing islands. We will also see an interesting dependence of these
two mechanisms on surface stress, that exists at free surfaces and also results
from lattice mismatch during heteroepitaxy.

Kinetic limitations might lead to either
two-dimensional growth or three-dimensional growth.
The former is achieved when atoms that land on islands can 
easily move down. 
However, when atoms which land on an island are hindered to
move down, islands nucleate on top of islands, and  a three-dimensional
structure results (see also 
Ref.~\cite{bet90}).  As in the above discussion of thermal equilibrium one
also often labels the kinetic growth modes according to the surface
morphologies of Fig.~\ref{growth_modes}, although the kinetic and the
thermodynamic limit represent totally different physics.

In order to gain insight into growth phenomena it is necessary 
to examine all possibly relevant microscopic processes
on the atomic scale. In Section~\ref{sec:atomistic} these
processes will be identified.  Section~\ref{sec:rates} then summarizes some
aspects of the description of growth by rate equations, and 
Section~\ref{sec:critical} analyses the conditions at which two-dimensional
growth is attained.

Theoretical methods that are promising for a reliable description of surface
diffusion and growth are described in Section~\ref{sec:3}.  In particular we
give a brief review of density functional theory in Section~\ref{sec:dft}.
Then, in Section~\ref{sec:implementation} we sketch how DFT is implemented in
actual computational schemes and utilized to identify microscopic processes
and to obtain growth-relevant parameters.  Section~\ref{sec:kmc} describes how these
{\em ab initio} parameters can be used to predict or analyze the temporal and
spatial evolution of epitaxial growth on macroscopic scales.  {\it Ab initio}
kinetic Monte Carlo simulations make the connection between the atomic scale
and time and length scales of realistic growth conditions.

Section~\ref{sec:results} then presents some recent results of close-packed
fcc surfaces.  We start with an analysis (but all these results are still
predictions) of the aluminum and silver (111) surfaces
(Sections~\ref{sec:al111} and 4.2).  Sections~\ref{sec:al100} and
\ref{sec:ag100} discuss results for aluminum and silver (100).
In particular, we consider the effects of stress for the two silver surfaces 
(Sections~\ref{sec:ag111str} and \ref{sec:ag100str}), possibly modified
by strain due to heteroepitaxy. For the Ag\,(111) surface we discuss in
Section~\ref{sec:surf} how adatom motion, the island density, and consequently the
growth mode can be influenced by surface active contaminants, so-called
surfactants.

\section{ATOMISTIC PROCESSES AND RATE EQUATIONS} 
\label{sec:2}

\subsection{Atomistic processes}
\label{sec:atomistic}

The conceptually simplest growth technique is {\it molecular beam epitaxy}
where an atom that has landed at the surface may either stay on the surface
and wander around, or evaporate back into the gas phase.  The latter happens
at a rate $\Gamma^{\rm ev} = \Gamma_0^{\rm ev} \exp(-E_{\rm ad}/{k_{\rm
    B}T})$, where $\Gamma_0^{\rm ev}$ is the effective attempt frequency (in
some works this term is approximated by the vibrational frequency of the
isolated adatom on the surface), $E_{\rm ad}$ is the adsorption energy of the
adatom, $k_{\rm B}$ the Boltzmann constant, and $T$ the substrate temperature
(for simplicity we assume here a situation of {\em atomic} desorption, i.e.,
no formation of molecules).  Typically $E_{\rm ad}$ is larger than the
activation barriers for other processes that occur on a surface so that
regardless of the exact magnitude of $\Gamma_0^{\rm ev}$ (typically of the
order of $10^{13}$s$^{-1}$) evaporation can be neglected during growth.

The different atomistic processes encountered by adatoms are illustrated in
Fig.~\ref{processes}.  After deposition $(a)$ atoms can diffuse across the
surface $(b)$ and will eventually meet another adatom to form a small nucleus
$(c)$ or get captured by an already existing island or a step edge $(d)$. Once
an adatom has been captured by an island, it may either break away from the
island ({\it reversible aggregation}) $(e)$ or remain bonded to the island
({\it irreversible aggregation}). An atom that is bonded to an island may
diffuse along its edge $(f)$ until it finds a favorable site.  As long as the
coverage of adsorbed material is low (say $\Theta \le 10~\%$), deposition on
top of islands is insignificant and nucleation of islands on top of existing
islands practically does not occur. However, if the step down motion $(g)$ is
hindered by an additional energy barrier, nucleation of island on top of
islands becomes likely $(h)$. 
\begin{figure}[tb]
\unitlength1cm
 \begin{center}
    \begin{picture}(10,6.5)
      \includegraphics{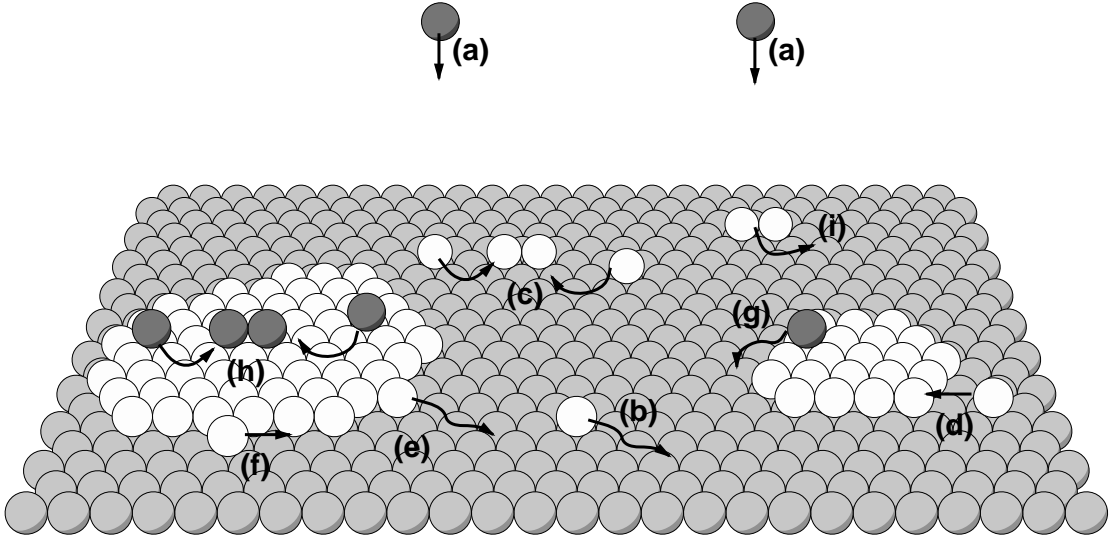}
    \end{picture}
 \end{center}
\caption{
The different atomistic processes for adatoms on a surface:
$(a)$ deposition,
$(b)$ diffusion at flat regions,
$(c)$ nucleation of an island,
$(d)$ diffusion towards and capture by a step edge,
$(e)$ detachment from an island, 
$(f)$ diffusion parallel to a step edge,
$(g)$ diffusion down from an upper to a lower terrace,
$(h)$ nucleation of an island on top of an already existing island, and
$(i)$ diffusion of a dimer (or a bigger island).
For the processes $(a), (c), (g)$ and $(h)$ also the reverse
direction is possible, but typically less likely.}
\label{processes}
\end{figure}

In principle it is possible that not just single adatoms but also dimers and
bigger islands migrate $(i)$.  For example, a dimer might diffuse by the two
atoms rotating around each other. Moreover, compared to a single adatom, a
dimer may be less bounded to the substrate since the electrons of the two
adatoms participate to the adatom-adatom bond and not only to the
adatom-substrate bonds. Therefore, it may be expected a low activation barrier
for the diffusion of dimers, but there is no clear evidence yet available.
Finally, it is sometimes believed that a large island is completely immobile.
However, results of Wen {\it et al.} \cite{Wen} for Ag/Ag\,(100) show that
even large scale clusters with $10^2$ to $10^3$ atoms can diffuse at room
temperature.  Diffusion of a cluster can either happen by consecutive edge
diffusion of single atoms from one side of the cluster to the other, or by
some concerted motion of all atoms in the cluster.  The importance of the
diffusion of dimers or large islands during growth is an issue that deserves
more attention in future research but will not be addressed any further in
this chapter.

Processes such as attachment to and detachment from step edges depend quite
sensitively on the local environment, because chemical bonding is a rather
local phenomenon and largely determined by the coordination of nearest
neighbor atoms. Figure~\ref{stepatoms} displays three important geometries of
step-edge atoms. In a bond cutting model of metallic bonding the energy of an
atom scales as the square root of the local coordination (see
Section~\ref{sec:glue} for more details).  In this approach it follows that
the binding energy of an atom at a kink site (atom 1 in Fig.~\ref{stepatoms})
equals the cohesive energy. This is indeed a general results and implies that
kink sites help to establish thermal equilibrium of the surface with the bulk
(note that a kink atom which detaches from a step creates a new kink in the
step edge).  Compared to an isolated adatom on the surface, the binding energy
of an atom in the step edge (atom 2 in Fig.~\ref{stepatoms}) is about 30~\%
and 60~\% larger than the binding energy of atom 1 and atom 3, respectively.
\begin{figure}[tb]
\unitlength1cm
\begin{center}
  \begin{picture}(10,4.2)
    \includegraphics{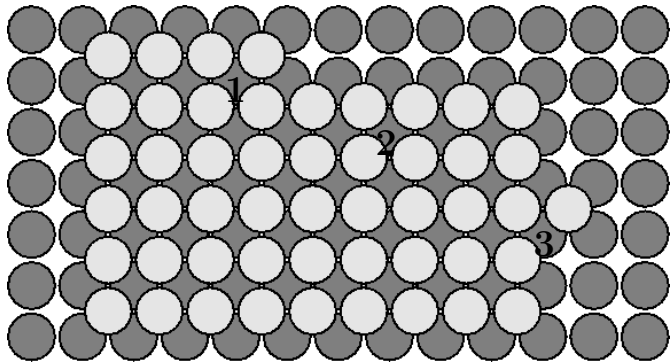}
    \thicklines
    \put(7.8,0.85){\bf 3}
    \put(5.7,2.2){\bf 2}
    \put(3.7,2.9){\bf 1}
  \end{picture}
\end{center}
\caption{
Three important geometries of atoms at step edges at an island on
a fcc\,(100) surface.}
\label{stepatoms}
\end{figure} 
The above discussion refers to metallic bonding and is
not necessarily valid for systems which can form covalent bonds and therefore
prefer a certain low coordination.

At not too high temperatures atoms will usually not detach from an island but
diffuse along the edge.  Eventually they will reach a higher coordinated site
such as kink site 1.  In general, fast edge diffusion leads to rather compact
island shapes, but when the edge diffusion is strongly hindered, adatoms
remain at the edge site where they reach the island and islands acquire a
fractal or dendritic form (``hit and stick mechanism''~\cite{san83}).

At a fcc\,(111) there are two close-packed steps (see Fig.~\ref{steps111}),
and because more open steps typically have a higher step formation energy,
these close-packed steps are expected to dominate the periphery of islands.
\begin{figure}[tb]
  \leavevmode \includegraphics{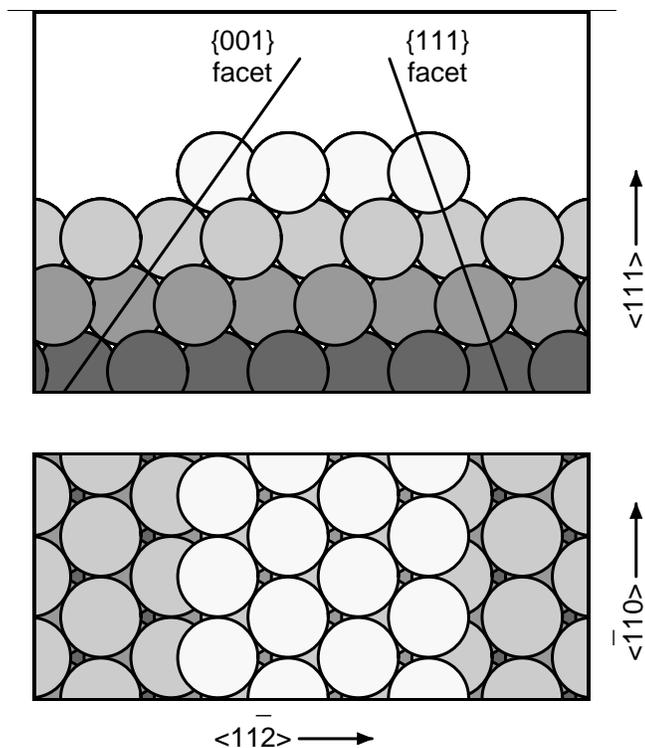} 
\vspace*{10cm}
\caption{
The two different types of close-packed steps on a fcc\,(111) surface.}
\label{steps111}
\end{figure} 
The steps are labeled as $\{100\}$ and $\{111\}$ facets,
referring to the plane passing through the atoms of the step and the atom of
the substrate (often these steps are labeled A and B).  Because of the
microscopic difference of the two types of steps the diffusion along them will
be different as well. It has been observed for growth of Pt on
Pt\,(111)~\cite{Michely1} that at a certain temperature the shape of the
islands observed is triangular. At a higher temperature, triangular islands
are observed again, but the triangles are rotated by $60^{\rm o}$. More
precisely, the islands are bounded only by $\{100\}$-faceted steps at a lower
temperature while at a higher temperature the islands are bounded by
$\{111\}$-faceted steps. It was proposed~\cite{Michely1} that this is a
consequence of the different diffusion constants for migration along the two
steps and particularly their different temperature dependences.

The key idea behind the kinetic description of the growth phenomena is that
processes occurring during growth, such as diffusion or desorption, are
described by rates.  The rate of a microscopic process $j$ that occurs during
growth usually has the form~\cite{gla41,vin57,wah90}
\begin{equation}
\Gamma^{(j)}= \frac{k_{\rm B}T}{h} \exp(-\Delta F^{(j)}/k_{\rm B}T)\quad ,
\label{ratdef}
\end{equation} 
where $\Delta F^{(j)}$ is the difference in the Helmholtz free
energy between the maximum (saddle point) and the minimum (equilibrium site)
of the potential curve along the reaction path of the process $j$. $T$ is the
temperature, $k_{\rm B}$ the Boltzmann constant, and $h$ the Planck constant.
The free energy of activation $\Delta F^{(j)}$ needed by the system to move
from the initial position to the saddle point is given by
\begin{equation}
\Delta F^{(j)} = E_{\rm d}^{(j)} - T \Delta S_{\rm vib}^{(j)}\quad .
\label{deltaf}
\end{equation} 
Here $E_{\rm d}^{(j)}$ is the sum of the differences in the
static total and vibrational energy of the system with the particle at the
minimum and at the saddle point, and $\Delta S_{\rm vib}^{(j)}$ is the
analogous difference in the vibrational entropy. The rate of the process $j$
can be cast as follows:
\begin{equation}
\Gamma^{(j)} = \Gamma_0^{(j)} \exp(-E_{\rm d}^{(j)}/k_{\rm B}T)\quad ,
\label{retdef1}
\end{equation} 
where $\Gamma_0^{(j)} = (k_{\rm B}T/{h}) \exp (\Delta S_{\rm
  vib}^{(j)}/k_{\rm B})$ is the effective attempt frequency. In the case of
isotropic motion of an adatom on the surface it follows from Eq.
(\ref{retdef1}) that the diffusion constant is $D = D_0 \exp(-E_{\rm
  d}^{(j)}/k_{\rm B}T)$~\cite{bob}. The prefactor $D_0 = 1/(2\alpha)
\Gamma_0^{(j)} l^2$ where $l$ is the jump length and $\alpha$ the
dimensionality of the motion ($\alpha =2$ for the surface).

The two basic quantities in Eq. (\ref{retdef1}) are the attempt frequency
$\Gamma_0^{(j)}$ and the activation energy $E_{\rm d}^{(j)}$.  Transition
state theory (TST)~\cite{vin57,wah90} allows an evaluation of $\Gamma_0^{(j)}$
within the harmonic approximation:
\begin{equation}
\Gamma_0^{\rm d} = \frac{\prod_{j = 1}^{3N} \, \nu_j}{\prod_{j = 1}^{3N-1} \,
  \nu_j^{*}}\quad ,
\label{harmappr}
\end{equation} 
where $\nu_j$ and $\nu_j^{*}$ are the normal mode frequencies
of the system with the adatom at the equilibrium site and at the saddle point,
respectively, and $3N$ is the number of degrees of freedom. The denominator in
Eq.~(\ref{harmappr}) contains the product of only $3N-1$ normal frequencies,
because for the adatom at the saddle point one of the mode describes the
motion of the particle toward the final site and has an imaginary frequency.
TST is only valid when $E_{\rm d}$ is larger than $k_{\rm B}T$.

The attempt frequency $\Gamma_0^{\rm d}$ shows a much weaker temperature
dependence than the exponential and for typical growth temperatures it is of
the order $10^{12} - 10^{13}s^{-1}$.  When the barriers for two different
diffusion events are different a {\it compensation effect} \cite{boi95} may
occur, i.e., $\Gamma_0^{\rm d}$ is larger for processes with a higher energy
barrier.  Indeed, a higher energy barrier usually implies a larger curvature
of the potential well around the equilibrium site of the adatom. The
corresponding vibrational frequencies of the adatom in such a potential are
larger as well, which implies [see Eq.~(\ref{harmappr})] that the attempt
frequency increases.

To define and determine $E_{\rm d}^{(j)}$ (and other quantities important for
the description of growth such as adsorption energies) we need to calculate
the ground-state total energy of the adsorbate system for a dense mesh of
adatom positions. This yields the so-called potential-energy surface (PES)
which is the potential energy experienced by the diffusing adatom,
\begin{equation}
E^{\rm PES}(X_{\rm ad}, Y_{\rm ad}) = \min_{Z_{\rm ad},\{{\bf R}_I\}}
E^{\rm tot} (X_{\rm ad}, Y_{\rm ad}, Z_{\rm ad},\{{\bf R}_I\}) \quad,
\label{PES}
\end{equation} 
where $E^{\rm tot} (X_{\rm ad}, Y_{\rm ad}, Z_{\rm ad},\{{\bf
  R}_I\})$ is the ground-state energy of the many-electron system (also
referred as the total energy) at the atomic configuration $(X_{\rm ad}, Y_{\rm
  ad}, Z_{\rm ad},\{{\bf R}_I\})$.  According to Eq.~(\ref{PES}) the PES is
the minimum of the total energy with respect to the $z$-coordinate of the
adatom $Z_{\rm ad}$ and all coordinates of the substrate atoms $\{{\bf
  R}_I\}$.  Assuming that vibrational effects can be neglected, the minima of
the PES represent stable and metastable sites of the adatom.  Note that this
PES refers to slow motion of nuclei and assumes that for any atomic
configuration the electrons are in their respective ground state. Thus, it is
assumed that the dynamics of the electrons and of the nuclei are decoupled.
This is the Born-Oppenheimer approximation that for not too high temperatures
is usually well justified.

\begin{figure}[tb]
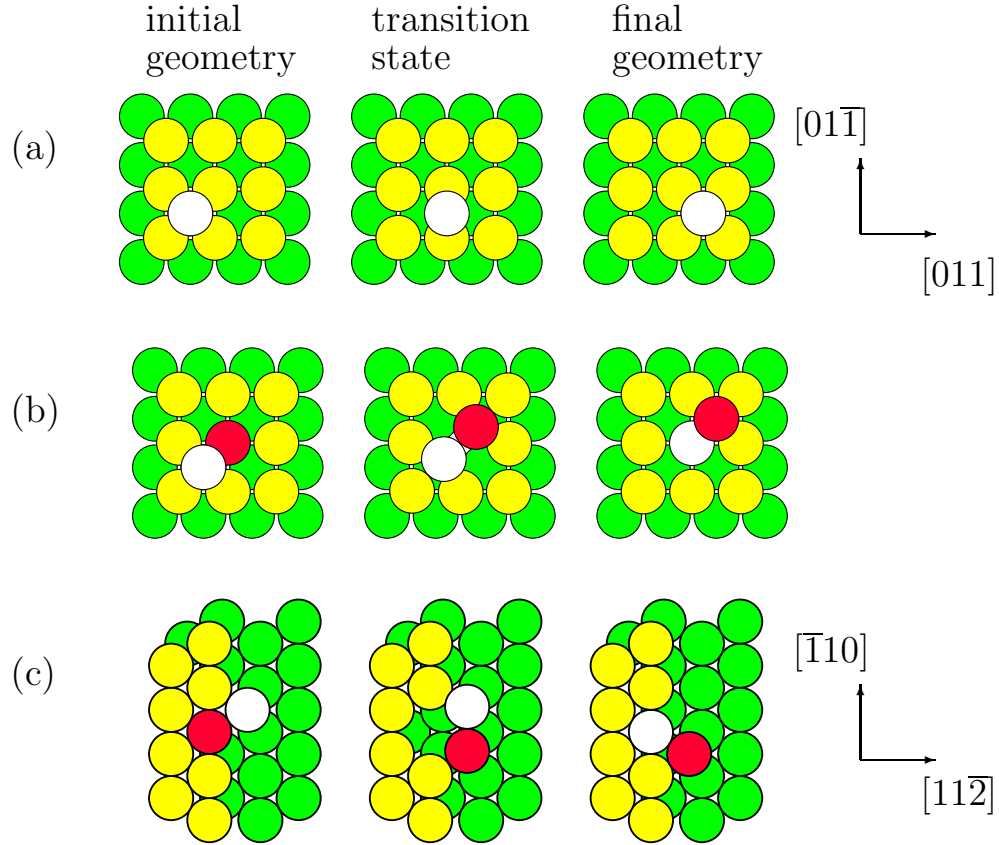

\unitlength1cm
\begin{center}
   \begin{picture}(10,10.0)
      \includegraphics{Fig5}
      \includegraphics{Fig6}
      \includegraphics{Fig7}
\put(-1.0,1){(c)}
\put(-1.0,4.5){(b)}
\put(-1.0,8){(a)}
\put(10.3,0.){\vector(0,1){1}}
\put(10.3,0.){\vector(1,0){1}}
\put(9.4,1.3){$[\overline{1}10]$}
\put(11.1,-0.6){$[11\overline{2}]$}
\put(10.3,7.0){\vector(0,1){1}}
\put(10.3,7.0){\vector(1,0){1}}
\put(9.4,8.3){$[01\overline{1}]$}
\put(11.1,6.3){$[011]$}
\put(0.8,9.7){{initial}}
\put(0.8,9.2){{geometry}}
\put(3.8,9.7){{transition}}
\put(3.8,9.2){{state}}
\put(7.0,9.7){{final}}
\put(7.0,9.2){{geometry}}
   \end{picture}
\end{center}
\caption{
Diffusion via hopping (a) and exchange (b) on a fcc\,(100) surface and 
diffusion along a \{111\} step on a fcc\,(111) surface via exchange (c).}
\label{more_exchange}
\end{figure}

Now consider all possible paths $l$ to get from one stable or metastable
adsorption site, ${\bf R}_{\rm ad}$, to an adjacent one, ${\bf R}_{\rm ad}{\bf
  '}$.  The energy difference $E_{{\rm d}l}$ between the energy at the saddle
point along $l$ and the energy of the initial geometry is the barrier for this
particular path. If the vibrational energy is negligible compared to $E_{{\rm
    d}l}$, the diffusion barrier then is the minimum value of all $E_{{\rm
    d}l}$ of the possible paths that connect ${\bf R}_{\rm ad}$ and ${\bf
  R}_{\rm ad}{\bf '}$, and the lowest energy saddle point is called the {\it
  transition state}. Although often only the path with the most favorable
energy barrier is important, it may happen that several paths exist with
comparable barriers or that the PES consists of more than one sheet (e.g.
Ref.~\cite{kley96}).  Then the {\it effective} barrier measured in an
experiment or a molecular dynamics (MD) simulation represents a proper average
over all possible pathways.

In the previous description it was assumed that an adatom moves from one
binding site to the nearest neighbor one. However, at higher temperatures
diffusing adatoms may from time to time jump over long distances, spanning
several lattice spacing~\cite{Ehrlich1}. Only little is known about this
process. In a recent experimental work on the diffusion of Pd on W\,(211)
Senft and Ehrlich~\cite{Ehrlich3} have extracted from their field ion
microscopy (FIM) measurements an activation barrier for long jumps roughly
twice that for single jumps. From the analysis of their experimental data they
have determined the temperature dependent probability for the occurrence of
very long jump (at least three nearest neighbor distances). These values
differ of at least one order of magnitude from the theoretical
ones~\cite{ferrando} and the reason for this discrepancy are still unknown.
More effort has to be put into a better understanding of the influence of such
long jumps on the intralayer transport.

But diffusion might also occur with a completely different mechanism, the
so-called {\em diffusion by atomic exchange} (or {\it exchange mechanism}).
The adatom can replace a surface atom and the replaced atom then assumes an
adsorption site. This was first discussed by Bassett and Webber~\cite{bas78}
and Wrigley and Ehrlich~\cite{wri78}. Even for the crystal bulk, namely Si,
exchange diffusion has been discussed~\cite{pand86}. This mechanism is
activated by the desire of the system to keep the number of cut bonds low
along the diffusion pathway.  On fcc\,(100) surfaces diffusion by atomic
exchange was observed and analyzed for Pt~\cite{kel90} and Ir~\cite{che90}.
For Al\,(100) it was predicted by Feibelman~\cite{exchange} and for Au\,(100)
by Yu and Scheffler~\cite{yu97}. The geometries for hopping and exchange
diffusion at a fcc\,(100) surface are shown in Figs.~\ref{more_exchange}(a)
and (b).

Diffusion along a step edge can also occur via the exchange mechanism as
illustrated in Fig.~\ref{more_exchange}(c) for a \{111\} step on the (111)
surface. An adatom at this step edge experiences a rather high diffusion
barrier, if the mechanism would be hopping: Either it has to move ontop of a
substrate atom or to leave the step edge to reach an adjacent step edge
position. However, if an {\it in-step} atom would move out of the step and the
adatom fill the opened site, the coordination of all the particles would not
decrease appreciably during the whole process.  Thus, the corresponding energy
barrier may be lower than that of the hopping process. It is also plausible that in
general the attempt frequency is different for a hopping and an exchange
process.

So far we have only discussed growth processes in the submonolayer regime. But
except when coverage is sufficiently low atoms might also land on top of an
existing island. Several questions arise at this point. If the island is large
enough and the adatom is far enough away from the edge of the island,
diffusion of this adatom usually will not differ from diffusion on the flat
terrace. However, this assumption is not always valid. The strain present on
the island may affect the self-diffusion barrier. Moreover, the atomic
structure ontop of the island may differ from the structure of the flat
surface. An example is given by Pt islands on Pt\,(111): at $T = 640$ K STM
images show reconstructed and unreconstructed terraces in coexistence with
different island densities~\cite{hoh95}. Furthermore, what happens if the
adatom is close to the island edge? Is the atom attracted by the edge?  Does
it stay on top of the island or is it hopping down?

It has been found first by Ehrlich and Hudda~\cite{Ehrlich2} and Schwoebel and
Shipsey~\cite{Schwoebel} and afterwards by a number of other studies that
metallic systems often exhibit an additional barrier hindering the diffusion
over a step edge as it is illustrated in Fig.~\ref{barrier}. This step-edge
barrier is often referred to as Ehrlich-Schwoebel barrier.
\begin{figure}[tb]
\input{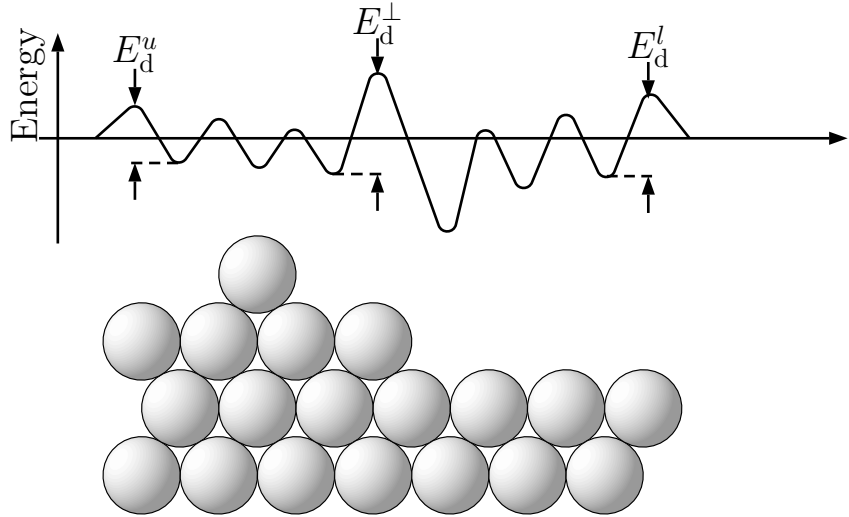}
\caption{
Schematic representation of the potential energy close to a step: 
$E_{\rm d}^{\perp}$ is the 
step-edge barrier, whereas $E_{\rm d}^{l}$ and $E_{\rm d}^{u}$ are the
diffusion barriers at the upper and lower terrace. The additional step-edge
barrier is $E_{\rm d}^{\perp} - E_{\rm d}^{u}$.}
\label{barrier}
\end{figure}
\begin{figure}[t]
\begin{center}
\input{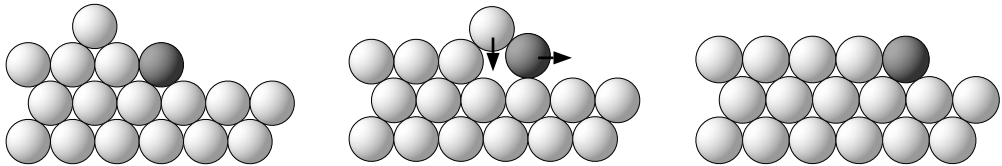}
\end{center}
\caption{
The motion of an atom from the upper terrace to the lower terrace  down 
by the exchange mechanism.}
\label{barrier2}
\end{figure} 
Intuitively its occurrence can be understood if hopping is the
relevant mechanism by employing simple bond counting arguments: The atom that
diffuses over the step edge is weaker bound right over the step edge because
at this position the number of bonds is reduced. This argument is valid for a
step down process by hopping.  The situation may be different in the case of
the exchange mechanism (cf.  Fig.~\ref{barrier2}) because here the number of
cut bonds remains low along the diffusion pathway. For some metallic systems
[for example, Al on Al\,(111) and Ag on Ag\,(100)] calculations have shown
that this is the favored situation (see Sections~\ref{sec:al111}
and~\ref{sec:ag100}). Note that the description of the process in terms of a
PES [see Eq. (\ref{PES})] valid for simple jumps of an adatom (i.e., diffusion
by hopping) holds for the diffusion by atomic exchange as well [see
Fig.~\ref{more_exchange}(b) and (c)].

The step-edge barrier determines in homoepitaxy whether the growth mode is
three-dimensional island growth or two-dimensional layer-by-layer growth. From
experiments this barrier has been estimated by analyzing STM
images~\cite{Meyer,bro95}. The idea is to measure the size of an island just
when nucleation on top of an island starts and to utilize nucleation theory to
estimate the step-edge barrier. Similar, indirect studies that interpret Monte
Carlo simulations \cite{Pavel1} will give an estimate of the step-edge
barrier. However, these approaches do not distinguish between different step
types and are unable to identify the microscopic mechanism for interlayer mass
transport.

\subsection{Rate equations}
\label{sec:rates}

Processes $(a), (b), (c), (d),$ and $(e)$ of Fig.~\ref{processes} form the basis of
phenomenological rate equations of the form
\begin{equation}
{dN_1 \over dt} = \Phi - 2\kappa_1 N_1^2 - N_1 \sum_{j>1} \kappa_j N_j
+ 2 \gamma_2 N_2 + \sum_{j>2} \gamma_j N_j
\label{RE1}
\end{equation}
\begin{equation}
{dN_j \over dt} = N_1 (\kappa_{j-1} N_{j-1} - \kappa_j N_j)
- \gamma_j N_j + \gamma_{j+1} N_{j+1} \quad .
\label{RE2}
\end{equation} 
These equations describe the time evolution of the adatom
density, $N_1$, and the density of islands of size $j$, $N_j$, for growth on a
flat surface in the submonolayer regime.  Adatoms are deposited onto the
substrate at a rate $\Phi = F \, \cal{N}$ adatom/s where $F$ is the flux in
ML/s and $\cal{N}$ the number of atoms pro ML.  The second and third term in
Eq.~(\ref{RE1}) account for isolated adatoms being ``lost'' because two
adatoms can meet at a rate $\kappa_1$ to form a new nucleus, or adatoms get
captured at a rate $\kappa_j$ by an island of size $j$.  The last two terms in
Eq.~(\ref{RE1}) describe further supply sources of adatoms and are gain terms
because dimers may dissociate and adatoms detach from an island of size $j$ at
a rate $\gamma_j$.  Equation~(\ref{RE2}) reflects the fact that the number of
islands of size $j$ increases because islands of size $j-1$ grow and islands
of size $j+1$ shrink. The number of islands of size $j$ decreases when islands
of size $j$ either shrink or grow.  Note that no evaporation into the gas
phase is included in Eqs.~(\ref{RE1}) and (\ref{RE2}) (that means that the
description in terms of Eqs.~(\ref{RE1}) and (\ref{RE2}) is appropriate only
at not too high temperatures). One could assume that the rate coefficients
$\kappa_j \propto D$ and $\gamma_j$ are independent on the island size and
surface coverage ({\it point island models}).  Another plausible choice is
that the rates depend on the length of the perimeter of the island so that a
first approximation is $\kappa_j \propto \gamma_j \propto \sqrt{j}$ for
compact islands. It has been shown~\cite{Bales} that the dependence on size
and coverage is more complex but this will not be discussed here.  The rate
coefficients $\kappa_s$ and $\gamma_s$ are only effective parameters and the
physics behind them remains unclear. For example, we will see below that the
different processes shown in Fig.~\ref{processes} may have different energy
barriers and different pre-exponential factors, and/or they may proceed by
different microscopic mechanisms (see Section~\ref{sec:results}).  In
principle, all these features can be taken into account in Eqs.~(\ref{RE1})
and (\ref{RE2}) by introducing different coefficients $\kappa_j$ and
$\gamma_j$ but they become less tractable and their clarity gets lost. A rate
equation analysis may help to gain some qualitative understanding of growth
processes but one can not expect insight into the {\it microscopic mechanisms}
governing growth.

With the assumption that agglomerates of $i^* +1$ and more adatoms are stable
against break-up ($\gamma_j = 0$ for $j > i^*$) one derives the scaling
relation~\cite{Scaling1}
\begin{equation}
N^{\rm is} \propto \left(D \over F\right)^{-{i^* \over i^* + 2}}
\label{scaling}
\end{equation} 
where $N^{\rm is}=\sum_{j>i^*}N_j$ is the island density and
$D$ the diffusion coefficient of an adatom on the flat surface.  The number
$i^*$ is called the size of the critical nucleus.  Relation (\ref{scaling})
can be used to extract microscopic parameters from experimental (in particular
STM) measurements: If one measures the island density as a function of $F$,
one can determine the critical nucleus size $i^*$.  With a known value for
$i^*$ (assuming that it does not change with temperature) one can determine
the diffusion barrier $E_{\rm d}$ and the prefactor $D_0$ if one measures the
temperature dependence of $N^{\rm is}$.  It is not the purpose of this article
to review scaling theory; however, we would like to point out that this method
is not as straightforward as often believed. For example, as pointed out by
Ratsch {\it et al.} \cite{PRL_Ratsch,SSL_Ratsch} the size of the critical
nucleus $i^*$ is not always well defined unless the temperature is
sufficiently low and $i^*=1$.

\subsection{Critical island area and the action of surfactants}
\label{sec:critical}

The definition of layer-by-layer growth mentioned at the beginning of
Section~\ref{sec:intro} trivially translates into the following
equation~\cite{sto82,Tersoff94,scheff96},
\begin{equation}
\label{eq:lbl}
A^{\rm is}_{c} \ge 1 / N^{\rm is} \quad ,
\end{equation} 
where $A^{\rm is}_{c}$ is the island area at which nucleation
sets in on top of an already existing island, and $N^{\rm is}$ is the island
density [cf. also Eq.~(\ref{scaling})].  As before and in all what follows we
assume that we are in a regime of island growth rather than step flow. If
Eq.~(\ref{eq:lbl}) is fulfilled, the islands will coalesce before a second
layer has started to grow.  The island density, $N^{\rm is}$, and the critical
island area, $A^{\rm is}_{c}$ are controlled by the growth conditions
(deposition rate and temperature) as well as by the different energy barriers
and interactions of the deposited adatoms on the surface and by the minimal
size of an island nucleus.  $A^{\rm is}_{c}$ is determined by the probability
that a number of $i^* + 1$ adatoms meet on the same island and form a stable
nucleus.

Without a step-edge barrier the adatoms that land on an island are not
hindered to move down and bind at the favorable sites. Thus, the formation of
a nucleus on top of an island becomes unlikely, and layer-by-layer growth is
expected.  However, the situation is different when atoms that land on top of
an island are hindered by a step-edge barrier to move downwards.  In this case
it is more likely that $i^* + 1$ adatoms meet to form a stable cluster that
subsequently will grow into a bigger and bigger island.  Thus, $A^{\rm is}_c$
is smaller (compared to $1/N^{\rm is})$ and it is more likely that islands on
the surface reach the area $A^{\rm is}_c$ before the layer is completed.  In
other words, when a noticeable step-edge barrier exists Eq.~(\ref{eq:lbl}) may
not be fulfilled and the system will grow three-dimensionally.

Eq.~(\ref{eq:lbl}) can be also rewritten in the following form
\begin{equation}
\label{eq:lbl2}
\Theta_{c} = A^{\rm is}_{c}  N^{\rm is} \ge 1  \quad ,
\end{equation} 
where $\Theta_{c}$ is the critical coverage for which, when it
is exceeded, islands will grow on top of already existing islands.  Looking at
experimental situations it appears that the conditions set by
Eqs.~(\ref{eq:lbl}) and (\ref{eq:lbl2}) are slightly too strong and it is
probably sufficient to request $\Theta_{c} \ge 0.9$ for good layer-by-layer
growth.  The importance of the above equations is that they show that the
growth mode can be influenced in two independent ways: One can modify the
island density (which is controlled by the adsorbate mobility at flat regions)
or one can modify the critical island area $A^{\rm is}_{c}$ which is largely
determined by the physics at step edges. Thus, it follows that the growth mode
can be changed from three-dimensional to layer-by-layer in the following ways:
\begin{itemize}
\item [1)] It has been shown by Kunkel {\em et al.} \cite{kunkel90} that on
  Pt\,(111) at very low temperatures the island shape is fractal. This implies
  that the island perimeter is particularly long, that the islands have rough
  edges, and many rather thin branches. As a consequence the step-edge barrier
  might be reduced substantially. But even if the step-edge barrier remained
  unchanged, the probability for an adatom which lands on such an island to
  move down to the lower terrace is high, because the adatom will visit the
  edge very frequently. This makes it likely that a possibly existing energy
  barrier can be overcome.  Moreover, a large number of kink sites or weakly
  bound atoms is present along the edges of the fractally shaped islands.
  Thus, an exchange downward diffusion of an adatom on the upper terrace at
  these sites may be more likely than the same mechanism involving an atom of
  a compact step edge.  Furthermore, we note that the island density $N^{\rm
    is}$ is high at low temperature [cf. Eq.~(\ref{scaling})] which reduces
  the probability that more than one atom land in a reasonable time interval
  on the same island.  Altogether these properties give rise to layer-by-layer
  growth.

\item [2)] A second possibility was demonstrated by Rosenfeld {\em et al.}
  \cite{rosenfeld93}. These authors have shown that increasing the island
  density $N^{\rm is}$ is in fact sufficient to achieve layer-by-layer growth.
  A high island density can be obtained for example by lowering the
  temperature or increasing the deposition rate in the very beginning of
  growth.  Once the island density is increased, the growth is continued at
  normal (higher $T$ and lower $F$) conditions.  Thus, the island density
  $N^{\rm is}$ is set large (to low $T$ and high $F$ parameters) but $A^{\rm
    is}_{c}$ is not reduced and remains at the value determined by the
  ``normal'' $T$ and $F$ parameters.  Indeed, three-dimensional growth did not
  start before the layer was completed.

\item [3)] For completeness we mention the possibility to enhance the mobility
  of deposited adatoms by photo-stimulation. However, we will not elaborate on
  this mechanism.

\item [4)] A very interesting way to achieve layer-by-layer growth uses
  surface contaminants, so called surfactants. There is one
  necessary condition these species should fulfill: They should stay on the
  surface during growth, thus they should not become buried during the
  growth process. While a good probability of surface segregation is necessary
  it alone would not affect the growth mode.  There are the following
  possible mechanisms that, when active, provide that a surface
  segregating contaminant increases the inter-layer mass transport:
\begin{itemize}
\item [$i)$] The simplest idea that comes to mind is that the surfactant
  decorates edges of steps and islands and reduces the step-edge barrier,
  since the atom-surfactant interaction is usually weaker than the atom-atom
  interaction.  Figure~\ref{barrier2} demonstrates how this could be achieved.
  Lowering the step edge barrier facilitates the interlayer transport and
  $\Theta_c \approx 1$. Oxygen for Pt\,(111)~\cite{esc94} and indium for
  Cu\,(100) seem to have this effect~\cite{veg95} and enhance the 2D character
  of the growth mode. Recently another picture has been proposed~\cite{mar94}:
  A repelling action of In at steps hampers the attachment of Cu atoms
  approaching the step edge from the lower terrace and gives rise to an
  enhancement of the island density.  However, the experimental data of
  Ref.~\cite{veg95} seems to rule out this scenario.
\item [$ii)$] It is also possible that surface impurities induce a potential 
  energy gradient that attracts deposited atoms towards the step;
  for deposited atoms that land on an island the number of visits at the edge 
  is thus increased, and as a consequence the probability to move down is 
  increased as well.
\item [$iii)$] Surfactants may act as nucleation centers, thus increasing the
  island density $N^{\rm is}$, while $A^{\rm is}_{c}$ remains unchanged (of
  course, an increase would be even better). This will induce layer-by-layer
  growth, provided that the probability that atoms which land on an island and
  move to the lower terrace is not reduced as well.  Moreover, if the
  surfactant increases the diffusion barriers $E_{\rm d}^{l}$ and $E_{\rm
    d}^{u}$ but leaves $E_{\rm d}^{\perp}$ (cf. Fig.~\ref{barrier})
  essentially unaffected, it follows that $E_{\rm d}^{\perp}-E_{\rm d}^{u}$ is
  reduced and the wanted effect may result.  This mechanism was recently
  discussed by Zhang and Lagally~\cite{zha94}.
\item [$iv)$] A forth possibility was discussed in the context of
  the surfactant action of Sb on Ag\,(111). The basic mechanism here is that
  Sb impurity atoms on the surface are practically immobile and
  act {\em repulsively} to deposited Ag adatoms.  This will also increase 
  the island density $N^{\rm is}$ and thus further two-dimensional growth.
  This mechanism will be discussed in Section~\ref{sec:surf}.
\end{itemize}
\end{itemize}

Tersoff {\em et al.} \cite{Tersoff94} have recently discussed
Eq.~(\ref{eq:lbl}) by assuming a circular island shape and various sizes for
the island nucleus.  They demonstrated that the critical island area is indeed
a useful concept.  However, we hesitate to give an explicit formula for it in
terms of diffusivities and deposition rate because in reality $A^{\rm is}_{c}$
will depend sensitively on the long- and medium-range adatom-adatom and
adatom-step interactions (see the attractive gradient towards the step on the
upper and lower terraces in Fig.~\ref{barrier}), as well as on the diffusion
barriers of adatoms parallel to step edges, as these determine the actual
shape of an island.

A study of Memmel and Bertel~\cite{mem95} has raised an interesting point.
They propose a simple model which connects the diffusion behavior on metal
surfaces to the charge density supplied by occupied two-dimensional
free-electron surface states. The argument is very appealing: A decrease in
the difference between the step edge barrier and the activation energy for
diffusion on the flat terraces could enhance the interlayer mass transport.
The barrier for the diffusion on the flat terrace is mainly determined by the
corrugation of the electron density to which both bulk Bloch states as well as
surface states contribute. The surface states are particularly interesting,
since they can strongly be influenced. A depopulation of these states induced
by confinement onto small islands or by the presence of an appropriate
surfactant increases the diffusion barrier on the flat surface with a
consequent reduction of the additional step edge barrier at the step edge.
Thus, an increased interlayer transport is expected with the related
layer-by-layer growth. This picture seems to be appropriate for the effects of
oxygen on the growth mode of Pt on Pt\,(111)~\cite{esc94}.

\section{TOTAL ENERGY AND THE DESCRIPTION OF GROWTH}
\label{sec:3}
In Section~\ref{sec:atomistic} we defined the potential energy surface (PES)
of a diffusing adatom. Obviously, the PES is governed by the breaking and
making of chemical bonds, and we also noted the need to take atomic
relaxations into account [cf. Eq. ({\ref{PES})].  Thus, the evaluation of the
PES requires an accurate, quantum-mechanical description of the
many-electron system. This can be achieved by modern density functional
theory calculations that combine electronic self-consistency and efficient
geometry optimization.  Approximate methods, based on the concepts of DFT
but employing empirical parameter instead of elaborate calculations have
been developed as well.  Such approximate methods are widely used by several
groups to investigate surface properties and to perform MD investigations of
adatom diffusion. We will sketch their main characteristics in
Section~\ref{sec:glue}.  A description of the basic concepts of DFT is then
given in Section~\ref{sec:dft}, and Section~\ref{sec:implementation}
describes how DFT is implemented into accurate self-consistent calculation
methods.  Finally, in Section~\ref{sec:kmc} we describe briefly the kinetic
Monte Carlo (KMC) technique that is capable to tackle the realistic time and
length scale of growth.

\subsection{Bond-cutting  methods}
\label{sec:glue}

Several methods have been developed based on the idea that the energy of a
many-electron, poly-atomic system can be written in terms of contributions
from the individual atoms:
\begin{equation}
E^{\rm tot}(\{ {\bf R}_I\}) = \sum_I E_I \quad .
\label{e-tot}
\end{equation} 
The sum goes over all atoms, and $E_I$ is the contribution of
the $I$-th atom. $E_I$ depends sensitively on the local geometry of atom $I$
(its embedding). The different bond-cutting methods differ in the treatment of
the actual form of the ``embedding function'' and in the way to determine the
necessary materials parameters.  The differences are not very significant and
the most popular names of these methods are: embedded atom method
(EAM)~\cite{daw83,vot94}, effective medium theory
(EMT)~\cite{nor80,sto80,jac87,nor90}, Finnis-Sinclair $N$-body
potentials~\cite{fin84}, second-moment approximation~\cite{duc71}, and
glue-model~\cite{erc86}.

In the simplest version of a bond-cutting approach it is assumed that the
energy per atom $E_I$ varies linearly with the atom's coordination number.
Thus, it is assumed that the strength of a bond is invariant of the number of
bonds the atom does form.  This approach clearly neglects the
quantum-mechanical properties of bonding, namely that the bond strength
saturates at a certain number of neighbors~\cite{duc71,rober91,met92}.  In
fact, detailed DFT studies have shown~\cite{rober91,met92} that the dependence
of $E_I$ on the local coordination is very similar to
\begin{equation}
E_I \approx - A  \sqrt{C_I} + B C_I \quad ,
\label{sqroot}
\end{equation}
with $C_{I}$ the coordination number of the ${I}$-th atom.

A more general approach gives
\begin{equation}
E_I =  F \left(\rho_I \right)
+ \frac{1}{2} \sum_{J \ne I} \phi ( |{\bf R}_J - {\bf R}_I| )  \quad .
\label{eameq}
\end{equation} Here $\phi$ describes the pair-wise, repulsive interaction
between atoms, and $F$ is called the {\it embedding function}, that depends on
the electron density created at site $I$.  For the effective medium theory
Christensen and Jacobsen~\cite{jac92} have shown that Eq.~({\ref{eameq})
resembles the behavior of the simple function noted in Eq.~(\ref{sqroot}),
but also contains some refinements.  Indeed, is has been shown that
Eqs.~(\ref{e-tot}) and (\ref{eameq}) represent an approximation of the
total-energy expression of density functional theory~\cite{jac87,nor90}.

The main problem in actual calculations is to determine the necessary
parameters to define the embedding function. Typically the parameters are
obtained by fitting results from a treatment based on Eq.~(\ref{eameq}) to
some experimental or DFT results of ``related systems''.  The results depend
on what systems and what properties are chosen.  The predictive power of these
methods has to be questioned (see e.g.~\cite{fei94,rat97}). We note, however,
that bond-cutting methods hold a significant share of the quantum-mechanical
description and thus are most valuable to summarize and to explain trends of
results obtained by DFT calculations.

\subsection{Density functional theory} 
\label{sec:dft}

The total energy of an $N^{\rm e}$-electron, poly-atomic system is given by
the expectation value of the many-particle Hamiltonian, using the many-body
wave-function of the electronic ground state.  For a solid or a surface the
calculation of such expectation value is impossible when using a wave-function
approach. However, as has been shown by Hohenberg and Kohn~\cite{hoh64}, the
ground-state total energy can also be obtained without explicit knowledge of
the many-electron wave-function, but from minimizing an energy functional
$E[n]$.  This is the essence of density functional theory (DFT), which is
primarily (though in principle not exclusively) a theory of the electronic
ground state, couched in terms of the electron density $n({\bf r})$ instead of
the many-electron wave function $\Psi(\{{\bf r}_i\})$ with ${\bf r}_i$ the
coordinates of the $i$-th electron.

The important theorem of Hohenberg and Kohn~\cite{hoh64} (see also
Levy~\cite{lev79}) tells: The specification of a ground state density $n({\bf
  r})$ determines the many-body wave function.  In other words, Hohenberg and
Kohn realized that for the ground state the known functional $n({\bf r}) = n[
\Psi] = \langle \Psi | \sum_i \delta({\bf r} - {\bf r}_i)| \Psi \rangle$ can
be inverted, i.e., $\Psi = \Psi[n({\bf r})]$. Although it was shown that
$\Psi[n]$ exists, its explicit form remains unknown.

With the help of this theorem the variational problem of the many-particle
Schr\"odinger equation transforms into a variational problem of an energy
functional:
\begin{equation}
E_0 \leq \langle \Psi|H^{\rm e}|\Psi \rangle = E_v[\Psi[n]] = E_v[n]\quad .
\label{variat}
\end{equation} 
Here $E_0$ is the energy of the ground state, $H^{\rm e}$ the
Hamiltonian of the electrons, $E_v[n] = \int d{\bf r}\, v^{\rm ext}({\bf r})
n({\bf r}) + G[n]$, and $v^{\rm ext}({\bf r})$ is the external potential
(typically $v^{\rm ext}({\bf r})$ is the Coulomb potential due to nuclei). In
this functional $n({\bf r})$ is the variable (the electron ground-state
density of any $N^{\rm e}$-electron system), and $v^{\rm ext}({\bf r})$ is
kept fixed.  $G[n]$ is a {\em universal} functional independent of the system,
i.e., independent of $v^{\rm ext}({\bf r})$.  For example, $G[n]$ is the same
for an H-atom, a CO-molecule, a solid etc.. The main advantage of this
approach is that $n({\bf r})$ only depends on three variables while
$\Psi(\{{\bf r}_i\})$ depends on many variables (the 3$N^{\rm e}$ coordinates
of all electrons)~\cite{spin}.  Thus, it is plausible that the variational
problem of $E_v[n]$ is easier to solve than that of $\langle \Psi|H^{\rm
  e}|\Psi \rangle$, yet the result for the ground-state energy and the ground
state electron density will be the same.  The total energy entering
Eq.~(\ref{PES}) is~\cite{units}
\begin{equation}
E^{\rm tot} ( \{{\bf R}_J \} ) = E_0( \{ {\bf R}_J \} ) +
\frac{1}{2}\sum_{J,J',J \ne J'} \frac{Z_J Z_{J'}}{|{\bf R}_J -
{\bf R}_{J'}|} \quad,
\label{tot}
\end{equation}
where $\{ {\bf R}_J\}$ includes all atoms, and $Z_J$ is the nuclear charge.

Earlier work (in particular the Thomas-Fermi approach) had shown that the
treatment of the kinetic energy $\langle \Psi | -\frac{1}{2}\nabla^2 | \Psi
\rangle$ is of particular importance and Kohn and Sham~\cite{koh65} therefore
wrote the energy functional in the form
\begin{equation} 
E_v[n]  =  T_s[n] + \int d{\bf r}\,  v^{\rm ext}({\bf r}) n({\bf r}) + 
\frac{1}{2}  \int d{\bf r}\,  v^{\rm H}({\bf r}) n({\bf r}) + E^{\rm xc}[n] \quad
,
\label{excdef}
\end{equation} 
where $T_s[n]$ is the functional of the kinetic energy of a
system of non-interacting electrons with density $n({\bf r})$, and $v^{\rm
  H}({\bf r}) = \int d{\bf r'} \frac{n({\bf r'})}{|{\bf r} - {\bf r'}|}$, the
Hartree potential, is the time-averaged electrostatic potential created by the
electron density, and $E^{\rm xc}[n]$ is the so-called exchange-correlation
functional. It accounts for the Pauli principle, dynamical correlations due to
the Coulomb repulsion, and the correction of the self-interaction included for
convenience in the Hartree term.

With Eq.~(\ref{excdef}) the problem of the unknown functional $G[n]$ is
transformed to one involving $T_s[n]$ and $E^{\rm xc}[n]$.  The functional
defined by Eq.~(\ref{excdef}) can be also modified by adding terms which
vanish at the correct electron density. Such a new functional $\tilde E_v[n]$
may converge faster towards the ground state or may depend less sensitive on
the input density. The latter implies that the input density does not need to
be very accurate, yet the resulting energy represents an acceptable
approximation for the correct total energy (see e.g. Ref.~\cite{hay97}).
Although the functional $T_s[n]$ is not known explicitly in a mathematically
closed form, it can be evaluated exactly by using the following ``detour''
proposed by Kohn and Sham. The variational principle applied to
Eq.~(\ref{excdef}) leads to
\begin{eqnarray}
\frac{\delta E_v[n]}{\delta n({\bf r})} & = &
\frac{\delta T_s[n]}{\delta n({\bf r})} + v^{\rm ext}({\bf r}) + v^{\rm H}({\bf
r}) +
\frac{\delta E^{\rm xc}[n]}{\delta n({\bf r}) }
\label{variational1} \\
& = & \frac{\delta T_s[n]}{\delta n({\bf r})} + v^{\rm eff}({\bf r}) = \mu
\quad ,
\label{variational2}
\end{eqnarray}
where $\mu$ is the Lagrange multiplier associated with the requirement
of a constant particle number and thus equals the electron chemical
potential. The effective potential is defined as
\begin{equation}
v^{\rm eff}({\bf r}) = v^{\rm ext}({\bf r}) + v^{\rm H}({\bf r}) +
v^{\rm xc}({\bf r}) \quad,
\label{eff_pot}
\end{equation} 
with $v^{\rm xc}({\bf r}) = \delta E^{\rm xc}[n] / \delta
n({\bf r})$, and $n({\bf r})$ is a ground-state density of any non-interacting
electron system, i.e.,
\begin{equation}
n({\bf r}) = \sum_{i=1} f_i | \phi_i({\bf r}) |^2 \quad,
\label{denans}
\end{equation} 
where we introduced the occupation numbers $f_i$ which is
simply the Fermi function. Because $T_s[n]$ is the kinetic-energy functional
of non-interacting electrons, Eq.~(\ref{variational2}) [together with
Eq.~(\ref{denans})] is solved by
\begin{equation}
\left[- \frac{1}{2} \nabla^2 + v^{\rm eff}({\bf r}) \right] \phi_i({\bf r}) =
\epsilon_i \phi_i({\bf r}) \quad.
\label{kseq}
\end{equation} 
These are the Kohn-Sham equations. They have to be solved
self-consistently together with Eqs.~(\ref{eff_pot}) and (\ref{denans}).  In
principle this gives the exact ground-state electron density and total energy
of a system of interacting electrons.

The above Eqs.~(\ref{excdef}) - (\ref{kseq}) contain one term which is not
known exactly. This is the exchange-correlation functional $E^{\rm xc}$.  For
a better understanding of this term it is instructive to introduce the
following Schr\"{o}dinger equation~\cite{jon89,har74,gun76}:
\begin{equation}
H_{\lambda}|\Psi_{\lambda}\rangle = \{ T + V_{\lambda} + 
\lambda
W \}
|\Psi_{\lambda}\rangle \quad ,
\label{harham}
\end{equation} 
where $\lambda$ defines the strength of the electron-electron
interaction: $\lambda$ = 0 corresponds to the noninteracting system and
$\lambda$ = 1 to the interacting physical system. $V_{\lambda} = \sum_i
v_{\lambda}({\bf r}_i)$ is an external potential chosen to maintain the ground
state density $n({\bf r})$ at its $\lambda$ = 1 value independently of
$\lambda$.  The exchange-correlation energy $E^{\rm xc}$ can be written in
a form which resembles a Coulomb interaction, 
\begin{equation}
E^{\rm xc}[n] 
= \frac{1}{2} 
\int \int d{\bf r} d{\bf r}'\, n({\bf r})
\frac{1}{|{\bf r} - {\bf r}'|} n^{\rm xc}({\bf r},{\bf r}') \quad ,
\label{nfexc}
\end{equation}
where $n^{\rm xc}({\bf r},{\bf r}')$ is called the exchange-correlation hole.
It can be written as
\begin{equation}
n^{\rm xc}({\bf r},{\bf r}') = n({\bf r}')[\tilde{g}({\bf r},{\bf r}') - 1]
\quad ,
\label{echdef}
\end{equation}
with the electron 
pair-correlation function $\tilde{g}({\bf r,r'})$.
Using a wave function formalism one then obtains~\cite{gun76}
\begin{equation}
\tilde{g}({\bf r},{\bf r}') = \int_0^1 d \lambda \, \left[ \frac{\langle\Psi_{\lambda} |
\hat{n}({\bf r})
\hat{n}({\bf r}')|\Psi_{\lambda}\rangle}
{n({\bf r})n({\bf r}')} - \frac{\delta({\bf r} - {\bf r}')}{n({\bf r})} 
\right] \quad ,
\label{pcfunc}
\end{equation}
where $\hat{n}$ is the density operator.
\begin{figure}[tb]
\input{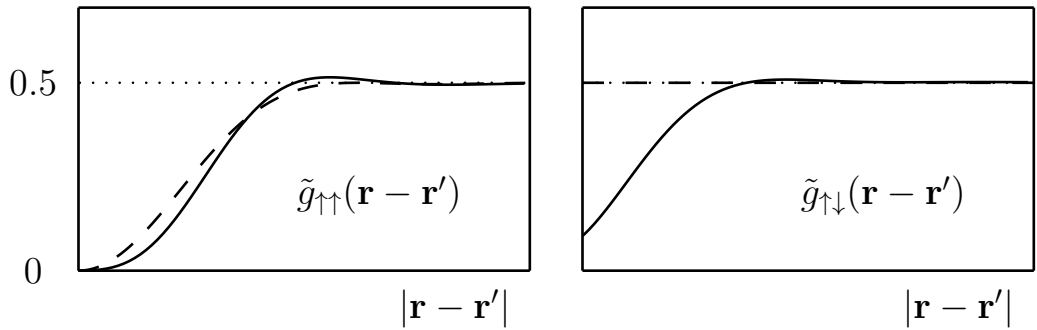}
\caption{
Sketch of the pair correlation functions $\tilde{g}_{\uparrow \downarrow}$ and
$\tilde{g}_{\uparrow \uparrow}$
  of a homogeneous electron gas in the Hartree (dotted
line) and Hartree-Fock (dashed line) approximation, and in the exact treatment 
(DFT) (solid line).}
\label{pair}
\end{figure} 
According to Eq.~(\ref{nfexc}), the exchange-correlation energy
arises from the Coulomb interaction of each electron (e.g. the one at ${\bf
  r}$) with a charge distribution $n^{\rm xc}({\bf r},{\bf r}')$, i.e., the
exchange-correlation hole surrounding that electron. The hole is a consequence
of the exchange and Coulomb interactions that cause a depletion of electron
density in the vicinity of each electron.  We note that the picture behind the
depletion is a dynamical one.  In the time average the depletion is not seen
in the electron density, but it gives rise to an important lowering of the
total energy, compared to that of a non-interacting electron system.

It is instructive to compare Hartree, Hartree-Fock, and DFT descriptions and
in Fig.~\ref{pair} we show schematically the spin-resolved pair-correlation
functions $\tilde{g}_{\uparrow \downarrow}$ and $\tilde{g}_{\uparrow
  \uparrow}$ of a many-electron system of constant density (also called a
jellium system).  The quantities $\tilde{g}_{\uparrow \downarrow}$ and
$\tilde{g}_{\uparrow \uparrow}$ are the pair-correlation functions for
electrons with parallel and antiparallel spins, respectively.  In a spin
unrestricted calculation $\tilde{g}$ of Eq.~(\ref{pcfunc}) is given by
$\tilde{g} = \tilde{g}_{\uparrow \downarrow} + \tilde{g}_{\uparrow \uparrow}$.
In the pure Hartree description exchange and correlation are ignored and the
pair-correlation functions are constant and equal to 1/2. The Hartree-Fock
method accounts for dynamical correlations due to the Pauli principle (the
exchange).  The pair-correlation function $\tilde{g}_{\uparrow \uparrow}$,
which is the probability of finding an electron at ${\bf r}'$ and an electron
at ${\bf r}$ with the same spin, shows a dependence on the interelectronic
spacing: The closer we are to the electron at ${\bf r}$, the lower is the
probability to find another electron.  However, $\tilde{g}_{\uparrow
  \downarrow}$ is constant because the Hartree-Fock description only contains
the ``Pauli correlation'' which affects electrons with the same spin, but the
depletion due to Coulomb repulsion which is independent of the spin is
neglected.  These are accounted for in DFT.  Thus, DFT gives a correct
description of the fact that electrons move in a correlated way and that this
correlation is due to the Pauli repulsion (for electrons of equal spin) and
the Coulomb repulsion (for all electrons).

The problem that remains in an actual DFT calculation is that the functional
$E^{\rm xc}[n]$ is unknown. However, some general properties of this
functional and values for some special cases are known.  Detailed and accurate
understanding exists for systems of constant electron density. The asymptotic
behavior at low and high densities is given by expressions derived by
Wigner~\cite{wig34} and Gell-Mann and Brueckner~\cite{gel57} and for
intermediate densities quantum Monte Carlo calculations have been performed by
Ceperley and Alder~\cite{cep80}. This gives the simple curve shown in
Fig.~\ref{excpar}, and this result for $\epsilon^{\rm xc}(r_s) = \epsilon^{\rm
  xc}_{\rm LDA}(n)$ is then used in the functional
\begin{equation}
E^{\rm xc}_{\rm LDA} [n] = \int d{\bf r}\, \, n({\bf r}) 
\epsilon^{\rm xc}_{\rm LDA} (n({\bf r}))\quad ,
\label{lda.exc}
\end{equation}
which is the local-density approximation (LDA)~\cite{koh65}.
Thus, in the LDA the many-body effects are included such that for
a homogeneous electron gas the treatment is exact and for
\begin{figure}[tb]
\unitlength1cm
\pspicture(10.0,7)   
\rput{-90}(8,3.6){\psfig{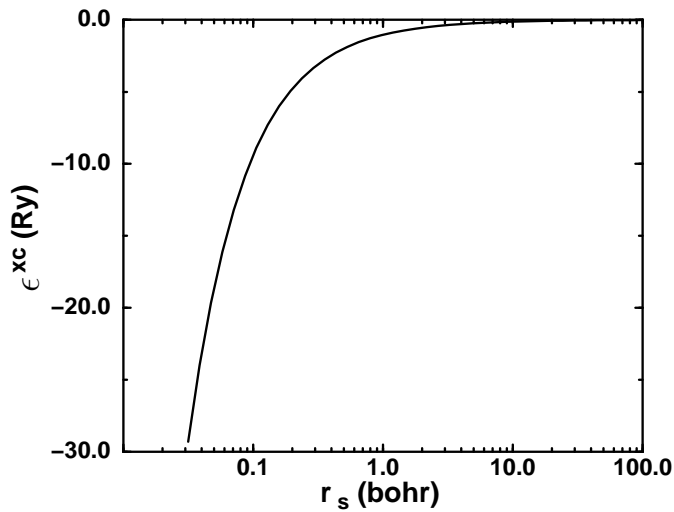}}
     \rput{90}(3.6,2.7){${\mathbf{\epsilon}}$}
\endpspicture
\caption{
Exchange-correlation energy per particle, $\epsilon^{\rm xc}$,
of homogeneous electron gases with density parameters $r_{\rm s}$. The electron
density and the density parameter are related by $n = \frac{4}{3} \pi r_s^3$.}
\label{excpar}
\end{figure} 
an inhomogeneous system exchange and correlation are treated by
assuming that the system is composed from many small systems with a locally
constant density.

The LDA can be improved by including the dependence on the density gradient
which leads to the generalized gradient approximation (GGA).  Several
different GGAs were proposed in the literature
\cite{per92,per86,heval,bec88,lee88,bec96} and have been used successfully for
DFT calculations of atoms, molecules, bulk solids, and surfaces (an overview
can be found in Refs.~\cite{bec96,per96}), but also limitations have been
identified for example by Mitas {\it et al.}~\cite{mit94} and Umrigar and
coworkers~\cite{umr96}.  It is by now clear that the lattice constants
calculated with a GGA are typically larger than those obtained with the LDA,
with the experimental values usually being in between.  Binding energies (or
cohesive energies) of molecules and solids as well as energy barriers of
chemical reactions are improved by the GGA (see Ref. \cite{ham94,ham95} and
references therein).  Still, for surface diffusion DFT-LDA calculations give
energy barriers in good agreement with those deduced from experiments and with
GGA calculations. Thus, although the total energies are changed when going
from the LDA to the GGA, the changes in energy barriers that are the {\it
  differences} between total energies are typically less pronounced (see e.g.
Ref.~\cite{rat97,yu96}).\\

\subsection{Implementation of DFT into state-of-the-art computations}
\label{sec:implementation}

Typically there are only a few possible 
adsorption sites and channels for diffusion. This is illustrated in 
Fig.~\ref{ad_sites}
\begin{figure}[tb]
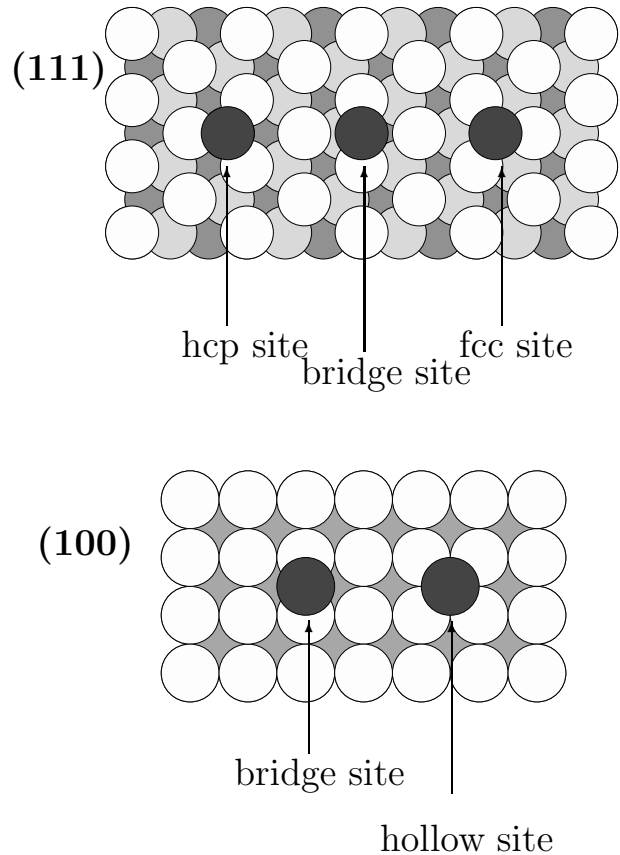

\leavevmode
\includegraphics{Fig12}
\includegraphics{Fig13}
\put(182,-133){\vector(0,1){60}}
\put(286,-133){\vector(0,1){60}}
\put(234,-143){\vector(0,1){70}}
\put(213,-295){\vector(0,1){50}}
\put(267,-310){\vector(0,1){65}}
\put(165,-145){hcp site}
\put(270,-145){fcc site}
\put(210,-155){bridge site}
\put(185,-307){bridge site}
\put(240,-332){hollow site}
\put(100,-40){\bf (111)}
\put(110,-220){\bf (100)}
\vspace*{1cm}
\caption{
Top view at a fcc\,(111) and (100) surface. The adsorption sites 
labeled fcc, hcp, and hollow site usually
correspond to the most stable binding sites 
while the bridge site is the transition state of a hopping diffusion.}
\label{ad_sites}
\end{figure} 
for the fcc\,(111) and fcc\,(100) surfaces.  For adatoms that are
chemically similar (or equivalent) to those of the substrate the stable sites
are those with high coordination and for hopping diffusion the transition
state is at the bridge site. The relevant information about the PES then is
obtained by calculating the total energy of the system with the adatom placed
in those positions.  In general, more care is necessary because the bridge
site could also be a local minimum of the PES and the energy barrier could be
in between the high coordination and the bridge sites. Furthermore, it is
possible that the diffusion does not proceed by hopping but by atomic
exchange~\cite{bas78,wri78,tun80,kel90,che90,exchange,yu97}.

In the bulk crystal the three-dimensional periodicity can be exploited by
using Bloch's theorem. Unfortunately, the presence of a surface and an adatom
on top of it breaks all periodicities. The (in principle) best approach to
treat such difficult situation is given by the Green-function
method~\cite{feibe1,schef1}.  An, at least in the past, popular approximation
for adsorbate systems is the cluster approach~\cite{Cluster}.  The presently
most efficient and practical approach that was also used in the results
discussed below is the supercell approach. The supercell may also be called a
big cluster, but in contrast to conventional cluster calculations the
supercell is periodically repeated.  As a consequence, the cluster boundary is
treated physically very accurately, and by utilizing the periodicity it is
possible to use very big cells.  The idea of an adatom on top of a substrate
in the supercell approach is sketched in Fig.~\ref{supercell}.
\begin{figure}[tb]
\leavevmode
\includegraphics{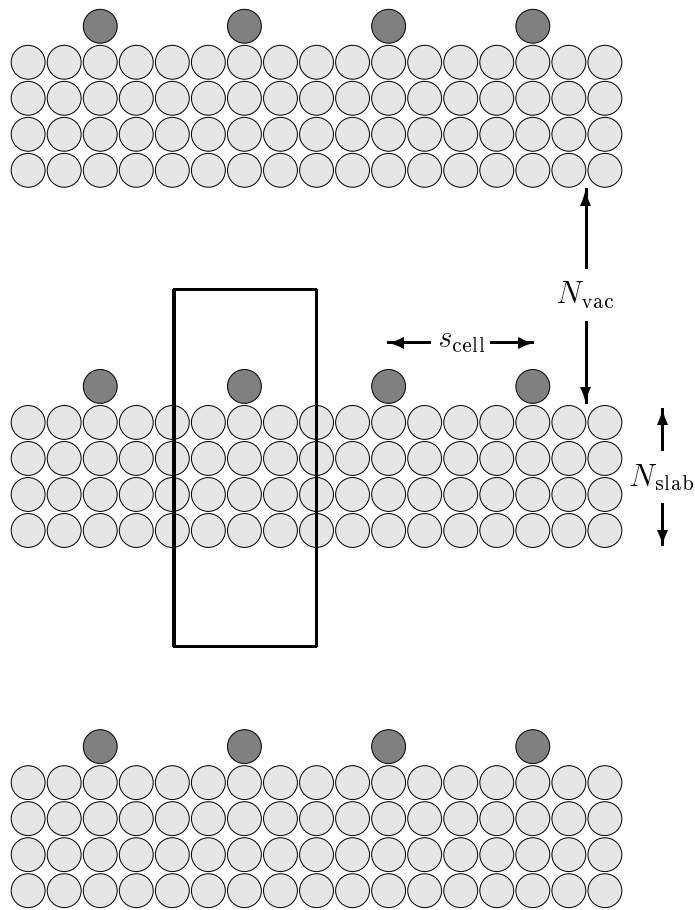}
\vspace*{13cm}
\caption{
Sketch of a supercell describing  an ``isolated'' adatom
at a surface (side view).}
\label{supercell}
\end{figure} 
The adatom is placed on top of a slab of a certain number of
layers.  The number of layers $N_{\rm slab}$ has to be sufficiently large so
that the adatom does not ``feel'' the presence of another surface on the other
side of the slab (or at least that the quantity to be computed, such as a
diffusion barrier, is not affected by the other surface).  Alternatively, one
could also place an adatom on either side of the slab; in this case, there are
more symmetries in the geometry but more layers are needed in the slab to
screen the mutual interaction between the two adatoms through the slab.  The
adequate number of layers $N_{\rm slab}$ depends on the properties that one
wants to calculate and the surface orientation, and careful tests have to be
carried out.  For example for the Ag\,(111) surface four layers are sufficient
when the adatom is placed on only one side of the slab, while for Al\,(100)
seven layers are necessary.

As illustrated in Fig.~\ref{supercell} the geometry repeats periodically in
vertical and lateral directions. The lateral periodicity implies that a single
adatom placed on a substrate is not at all a single adatom; if the cell size
parallel to the surface is chosen for example as ($2 \times 2$) we actually
calculate a system with a coverage of $25~\%$. It is therefore important to
test that the interaction with the neighboring adatoms can be neglected. On a
(111) surface a cell size of ($2 \times 2$) is usually sufficient, but
sometimes larger cells [($3 \times 3$) or even ($4 \times 4$)] are necessary.
To model a diffusion event along or across a step one either chooses a small
island on top of a substrate or a vicinal surface. The latter has the
advantage that there is only one step in the unit cell so that a smaller cell
size is required in order to attain a negligible step-step interaction. The
system also repeats in vertical direction separated by a vacuum region.  The
thickness of the vacuum region has to be tested as well, but the computational
cost of a thicker vacuum region is relatively small compared to a larger cell
size or a higher number of slab layers (for a deeper discussion of the above
technicalities see e.g. Ref.~\cite{stu96}).

Core electrons typically do not take part directly in the binding process of
atoms in molecules and solids, and the nature of the chemical bond is mainly
determined by the valence electrons. This is exploited by the {\it frozen core
  approximation} where the core electrons are effectively combined with the
nuclei to form frozen unpolarizable ions.  Still, not just the electrostatic
potential but also the quantum nature of the core electrons is felt by the
valence electrons. For example, different wave functions have to be orthogonal
and therefore the valence wave functions have nodes and oscillate in the core
region.  For practical calculations one needs to expand the wave function in a
suitable basis and we choose a plane wave basis set ~\cite{ihm79}
\begin{equation}
|\phi_j({\bf k,r})\rangle =  \sum_{{\bf G}}
c_{j,{\bf k}}({\bf G}) |{\bf G+k} \rangle \quad .
\label{plane_wave}
\end{equation} 
A plane wave description of wave functions that have nodes and
oscillate requires a very large number of plane waves. This inconvenience is
cured efficiently by the {\it pseudopotential} approach.  Modern {\em ab
  initio} pseudopotentials reproduce the potential of an atom exactly outside
the core region defined by a radius $r_c$ and are rather smooth inside the
core region.  An important requirement on a ``good'' pseudopotential is that
it is transferable. This means that the pseudopotential should behave like the
all-electron potential in a variety of different chemical situations.
Pseudopotentials that reproduce the same charge inside the core region as the
all-electron potential, and therefore have the same scattering properties, are
referred to as {\it norm-conserving}.  Those that are often used have been
developed by Bachelet, Hamann, and Schl\"{u}ter \cite{Bachelet}, Troullier and
Martins \cite{Troullier}, and Gonze, Stumpf, and Scheffler~\cite{gss}.
Recently, Vanderbilt \cite{Vanderbilt} proposed {\em ab initio}
pseudopotentials that drop the condition of norm-conservation and therefore
can be used with a lower number of plane waves. The gain in computer time due
to the smaller basis set is partially compensated by the costs to calculate
the correction required by the neglected norm-conservation.

The electron density is calculated according to Eq.~(\ref{denans}) as
\begin{equation}
n({\bf r}) = \sum_{\bf k} \sum_j \omega_{\bf k} f(\epsilon_j({\bf k})) |\phi_j({\bf k,r})|^2
\label{ksum}
\end{equation} 
where the integration over the Brillouin zone has been replaced
by a sum over a mesh of ${\bf k}$-points and $\omega_{\bf k}$ is the weight of
the ${\bf k}$-point.  A convenient scheme to construct an appropriate {\bf
  k}-point mesh is described by Monkhorst and Pack~\cite{Monkhorst}.  In {\em
  ab initio} pseudopotential calculations some matrix elements and some
integrals are efficiently evaluated in real space whereas others are
efficiently evaluated in reciprocal space. The technique of fast Fourier
transformation enables a numerically fast change from one representation to
the other.  Technical details of the computational procedures are described
for example in~\cite{Payne,kohler96,bock97}.

\subsection{Kinetic Monte Carlo approach}
\label{sec:kmc}

MD simulations can provide important insight into elementary microscopic
mechanisms but typically they cannot be used for growth studies.  The time
between two successful diffusion events is often of the order of nanoseconds.
During this time the adatoms undergo several (e.g.  $10^3$ and more)
unsuccessful attempts. Since MD calculates all these unsuccessful atomic
movements explicitly, MD simulations can cover times of some picoseconds,
possibly some nanoseconds and therefore are usually inappropriate to describe
the spatial and temporal evolution of growth patterns, that typically develop
on a time scale of seconds.  Instead, the method of choice for studying the
spatial and temporal development of growth is kinetic Monte Carlo
(KMC)~\cite{MC_references2}.  The key idea behind KMC is to describe
stochastic processes (such as deposition, diffusion, desorption, etc.) on the
microscopic scales by rates $\Gamma^{(j)} = \Gamma_0^{(j)} \exp(-E_{\rm
  d}^{(j)}/k_{\rm B}T)$, that were discussed in Section~\ref{sec:atomistic}
above.  In KMC a process $j$ is thermally activated with a relative
probability given by:
\begin{equation}
 w^{j} = \frac{\Gamma^{(j)}}{R} = \frac{\Gamma_0^{(j)}}{R} \, \exp(-E_{\rm d}^{(j)}/k_{\rm   B}T) \quad .
\label{probabi}
\end{equation} 
We define the total rate $R = \sum_j \Gamma^{(j)}$, where the
sum runs over all possible processes. Deposition is accounted for in this
description by the deposition rate $\Gamma_0^{(0)}=F$.  Equation
(\ref{probabi}) satisfies the condition of detailed balance.  Using a
stochastic approach the explicit calculation of unsuccessful attempts is
avoided.  Yet, the result of a KMC study will be the same as that of a MD
simulation, provided that the underlying PES is the same.  The strategy of KMC
can be summarized as follows:
\begin{enumerate}
\item[$1)$] Determination of all processes $j$ that possibly could take place with
  the actual configuration of the system.
\item[$2)$] Calculation of $R$.
\item[$3)$] Choose two random numbers $\rho_1$ and $\rho_2$ in the range
$(0,1]$.
\item[$4)$] Find the integer number $l$ for which
\begin{equation}
\sum_{j = 0}^{l - 1} \Gamma^{(j)} \leq \rho_1 \, R < \sum_{j = 0}^{l}
\Gamma^{(j)}\quad .
\label{mcalgo}
\end{equation}
\item[$5)$] Execute process $l$.
\item[$6)$] Update the simulation time $t := t + \Delta t$ with 
$\Delta t= -ln(\rho_2)/R\quad .$
\item[$7)$] Go back to step $1)$.
\end{enumerate}
Step $6)$ ensures that a direct and 
unambiguous relationship between KMC time and real time is established, 
since the KMC algorithm effectively simulates Poisson processes.

KMC differs from the algorithm proposed by Metropolis {\it et
  al.}~\cite{met53} that was successfully employed to determine the static
properties of many-particle systems. In the Metropolis Monte Carlo scheme the
probability that a new configuration is accepted is proportional to
$\exp(-\Delta E/k_{\rm B}{\rm T})$ where $\Delta E$ is the difference between
the total energies of the system in the new and old configuration. Such an
algorithm searches for the configuration corresponding to the minimum of the
total energy, and the sequence of generated configurations does not correspond
to the real time evolution of the system.

KMC simulations have been used to study crystal growth of semiconductors
(e.g.~\cite{semimc1,semimc2,semimc3}) and metals
(e.g.~\cite{work1,work2,work3,work4}).  However, most of these studies have
been based on restrictive approximations.  For example, the input parameters
have been treated as effective parameters determined rather indirectly by
fitting to experimental quantities, like intensity oscillations in helium atom
scattering (HAS) measurements or reflection high energy electron diffraction
(RHEED), or they were obtained from STM studies of island densities. Thus, the
connection between these parameters and the microscopic nature of the
processes may be somewhat uncertain.  Often the correct surface structure was
neglected and the simulation was done on a simple cubic lattice while the
system of interest had an fcc or bcc structure.  Despite these approximations
such studies have provided qualitative and in some cases also quantitative
insight into growth phenomena. It is desirable to carry out KMC simulations
with the proper geometry and microscopically well founded parameters. This has
been done for example in Refs.~\cite{liu93,jac95,sh96} where semi-empirical
calculations such as the embedded atom method or effective medium theory (cf.
Section~\ref{sec:glue}) have been employed to evaluate the PES. However, the
most accurate, but also most elaborate approach to obtain the PES employs DFT
as described in Sections~\ref{sec:dft} and \ref{sec:implementation}. Results
of a KMC study performed with input from such DFT calculations are presented
in Section~\ref{sec:kmcal}.

\section{RESULTS FOR FCC\,(111) AND FCC\,(100) SURFACES}
\label{sec:results}
We now discuss some recent DFT studies of various growth phenomena at (111)
and (100) surfaces of fcc metals, in particular aluminum
(Sections~\ref{sec:al111} and \ref{sec:al100}) and silver (Sections 4.2 and
\ref{sec:ag100}).  Section~\ref{sec:al111mic} presents DFT calculations for
the self-diffusion at Al\,(111), and these results form the basis of the {\em
  ab initio} KMC simulations described in Section~\ref{sec:kmcal}.
Section~\ref{sec:al100} summarizes the understanding of self-diffusion at
Al\,(100), and in Section~\ref{sec:ag100} we identify the microscopic
processes on the Ag\,(100) surface.

Surface stress can affect diffusion and crystal growth quite substantially.
Experimental investigations have been carried out for the self-diffusion at
Ag\,(111) and at thin silver films on Pt\,(111)~\cite{ker95}.  In
Section~\ref{sec:ag111str} we present results of recent DFT calculations for
these systems.  Section~\ref{sec:ag100str} discusses the influence of surface
stress on the diffusion mechanism, showing that stress can even change the
mechanisms.
 
One goal of epitaxial growth is to achieve atomically flat and defect free
surfaces of specified crystallographic orientation under convenient growth
conditions, i.e., not too low deposition rates and not too high temperatures.
In many cases this goal can be reached only with the use of surface
contaminants that act as surfactants~\cite{sny93,veg92}.
Section~\ref{sec:surf} presents DFT results concerned with the surfactant
functioning of Sb for the growth of Ag\,(111).

\subsection{Growth at Al\,(111)}
\label{sec:al111}
\subsubsection{Microscopic processes}  
\label{sec:al111mic}
On the (111) surface of a fcc crystal there are two kinds of close packed
steps, that were already discussed (see Fig.~\ref{steps111}).  Although the
packing of the step-edge atoms is identical in these two steps, the
coordination of the substrate atoms at the step edge is different: At the
$\{111\}$-faceted step the substrate atoms have a coordination of 11, and at
the $\{100\}$-faceted step the substrate atom coordination is 10.  Thus, a
small difference in the step-formation energy is expected. If lattice
relaxations are negligible, the $\{111\}$-faceted step might be slightly
favored.  Indeed, DFT calculations~\cite{stu96,stu94} confirm the trend
suggested by the coordination numbers: The $\{111\}$ facet is slightly favored
with $\sigma^{\{111\}} = 0.232$ eV per atom over the $\{100\}$ with
$\sigma^{\{100\}} = 0.248 $ eV per atom. Neglecting contributions from more
open steps, as these have a higher formation energy, one can now predict the
equilibrium shape of islands.  According to the Wulff construction in
thermodynamic equilibrium the island borders have distances from the island
center that are proportional to their formation energies.  As a consequence
the equilibrium shape of islands for Al/Al\,(111) should be hexagonal with the
edges alternating between those with a shorter $\{100\}$ and a longer
$\{111\}$ facet. The resulting ratio of the lengths $L^{\{100\}}/L^{\{111\}}$
is $4:5$.
\begin{figure}[b]
\leavevmode
\includegraphics{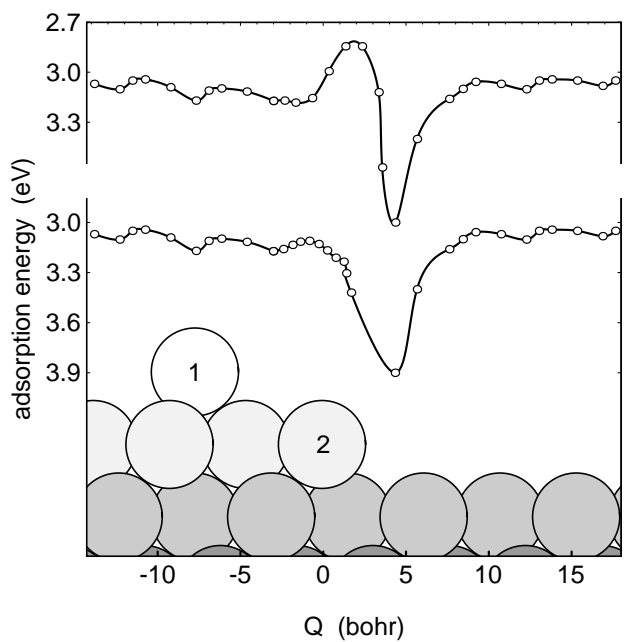}
\vspace*{8cm}
\caption{
Total energy along the diffusion path of an Al adatom over a
  $\{111\}$-faceted step on Al(111). The upper curve is evaluated for a
  hopping process, while the lower one refers to an exchange process. 
The generalized coordinate is 
$Q = X_1 + X_2$ with $X_I$ = $x$-coordinate of the adatom $I$.
The $x$ axis is parallel to the surface and
perpendicular to the step orientation. For the undistorted step $X_2 = 0$.}
\label{en.cur.111}
\end{figure} 

The activation barriers for the most important processes of self-diffusion at
Al\,(100) are collected in Table~\ref{diff.en.111}.  In addition we note that
the calculations predict that an isolated adatom on the flat surface favors
the hcp site (see Fig.~\ref{ad_sites}) and that the energies of the bridge and
fcc positions are almost degenerate.
\begin{table}[b]
\caption{
Energy barriers $E_{\rm d}$ for different self-diffusion processes on Al\,(111).}
\label{diff.en.111}
\begin{tabular}{lcccc}
\hline
process & \hspace{2.5cm} & mechanism & \hspace{2.5cm} & $E_{\rm d}$ (eV) \\
\hline
 flat Al(111) & & hopping & & $ 0.04 $ \\
 $\{111\}$ step $\|$ & & exchange & & $ 0.42 $ \\
 $\{111\}$ step $\|$ & & hopping & & $ 0.48 $ \\
 $\{100\}$ step $\|$ & & hopping & & $ 0.32 $ \\
 $\{100\}$ step $\|$ & & exchange & & $ 0.44 $ \\
 $\{111\}$ step $\perp$ descent & & exchange & & $ 0.06 $ \\
 $\{111\}$ step $\perp$ descent & & hopping & & $ 0.33 $ \\
 $\{100\}$ step $\perp$ descent & & exchange & & $ 0.08 $ \\
 $\{100\}$ step $\perp$ descent & & hopping & & $0.45 $ \\
\hline
\end{tabular}
\end{table} 
Therefore, the diffusion path for an isolated adatom goes from one
hcp site to the adjacent one through fcc and bridge positions. Since the fcc
site continues the ABCABC stacking of the fcc crystal, whereas the equally
threefold coordinated hcp position belongs to an ABCAC stacking, this result
is somewhat unexpected.  The epitaxial continuation of the crystal with
adatoms occupying fcc sites is recovered when the coverage is increased
($\Theta \ge 1/4$ ML).

The DFT calculations also predict a long-range attraction of adatoms towards
step edges for approach from the upper as well as from the lower terrace.  It
appears that this attraction is actuated by electronic surface states.  The
attraction is weak at long distances but close to the step it becomes so
strong that particularly at the lower terrace an adatom will be funneled
toward the step. This is clearly visible in Fig.~\ref{en.cur.111} (lower
curve) where the total energy along the adatom diffusion path involving the
migration toward and over a step and on the flat surface is displayed for the
hopping (upper curve) and the exchange (lower curve) mechanism. The attraction
is present with and without relaxation of the system and thus cannot be
elastic.  An electrostatic origin can also be discarded, since the dipoles
located at the step and of the isolated adatom have the same sign. Thus, the
resulting dipole-dipole interaction is repulsive. However, the adatom and step
induce localized electronic states that interact and it was
concluded~\cite{stu96,stu94} that they are responsible for the long-range
attractive interaction.

Figure~\ref{en.cur.111} also shows the difference between the hopping and the
exchange mechanism for the diffusion across the step from the higher to the
lower terrace. The upper curve is calculated for the hopping process, and the
presence of an energy barrier that hinders the roll over of the adatom from
the upper to the lower terrace is clearly seen.  On the other hand,
practically no hindrance exists for diffusion by exchange (cf.
Fig.~\ref{barrier2}).  The activation barrier for the exchange process for the
step-down motion is very low and comparable to the diffusion barrier on the
flat surface: 0.08 eV and 0.06 eV for $\{100\}$- and $\{111\}$-faceted steps,
respectively, whereas it is 0.04 eV for diffusion on the flat surface.  Thus,
we predict layer-by-layer growth for Al on Al\,(111) for a wide range of
substrate temperatures. The barriers for the exchange might hamper 2D growth
only at $T \approx 25$ K. However, at such a low temperature the island edges
are frayed which may reduce the barriers resulting in layer-by-layer growth.

One origin for the preference of the exchange process at step edges might be
the bonding character of Al.  Although Al is often considered as a
jellium-like metal, it is a rather covalent atom with its $sp$ valence
electrons (we remind that Al and As form the AlAs compound, a covalent,
zincblende semiconductor).  Thus, similarly as discussed by
Pandey~\cite{pand86} for the exchange diffusion in Si bulk, and by
Feibelman~\cite{exchange} for exchange diffusion at Al\,(100), we believe that
this mechanism is favored by the tendency of the system to keep a low number
of cut bonds along the diffusion pathway.  We expect that exchange diffusion
is a rather common mechanism for down movement of adatoms at step edges.

We now address the diffusion of adatoms parallel to the two close-packed
steps. As will be shown in Section~\ref{sec:kmcal}, this is of particular
importance for the shapes of islands which develop during growth.  The
migration along the steps may take place via a hopping or an exchange
mechanism.  The calculations predict that along the $\{100\}$-faceted step an
adatom preferentially jumps to an adjacent site with an activation barrier of
0.32 eV, whereas along the $\{111\}$ facet diffusion by atomic exchange is
preferred with activation barrier of 0.42 eV.  To understand this difference
we consider the positions labeled TS$_{\{100\}}$ and TS$_{\{111\}}$ in
Fig.~\ref{lonbrid} that are the lowest-energy transition state of a hop along
both steps.
\begin{figure}[tb]
\input{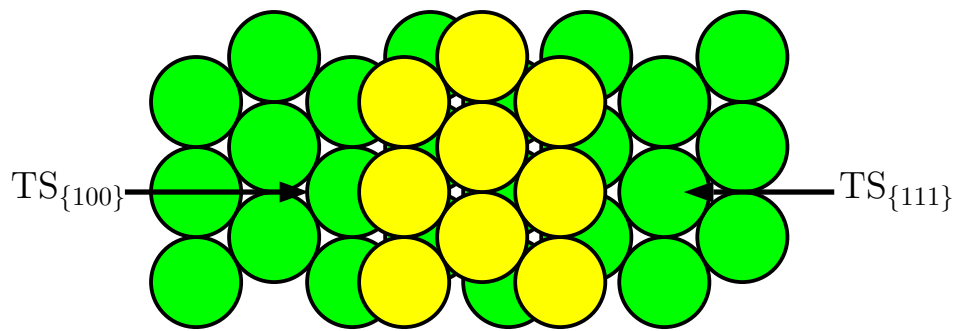}
\caption{
The transition sites for the hopping diffusion along the two
  close-packed steps on Al\,(111).}
\label{lonbrid}
\end{figure} 
Along the $\{111\}$-faceted step the adatom in the transition
state has three neighbors, whereas in the same position along the
$\{100\}$-faceted step the adatom has four neighbors~\cite{stu96}. The higher
coordination suggests a lower barrier for hopping along the $\{100\}$-faceted
step.  Indeed, the calculations yield a value of 0.32 eV, smaller than the
value of 0.48 eV obtained for the hopping along the $\{111\}$-faceted step.
The calculated barrier for the exchange mechanism is about the same for both
steps (0.42 and 0.44 eV).

\subsubsection{Ab initio KMC study of growth}    
\label{sec:kmcal}

We now analyze typical growth conditions where kinetic processes are dominant.
The detailed characterization of the energetics of diffusion processes carried
out by Stumpf and Scheffler~\cite{stu96,stu94} for Al/Al\,(111) and presented
in the previous Section has provided several parameters for realistic KMC
simulations.  Among the processes listed in Table~\ref{diff.en.111} we have
considered the following diffusion mechanisms:
\begin{enumerate}
  \item[$(i)$] diffusion of a single adatom on the flat 
    surface: $E_{\rm d}$ = 0.04 eV;
  \item[$(ii)$] exchange diffusion from upper to lower terraces:
    $E_{\rm d}$ = 0.06 eV at the \{100\}-faceted step and 
    $E_{\rm d}$ = 0.08 eV at the \{111\}-faceted step;
  \item[$(iii)$] diffusion parallel to the \{100\}-faceted step via 
    hopping: $E_{\rm d}$ = 0.32 eV;
  \item[$(iv)$] diffusion parallel to the \{111\}-faceted step via 
    exchange: $E_{\rm d}$ = 0.42 eV.
\end{enumerate}

The DFT calculations give that the binding energy of a dimer is 0.58
eV~\cite{stu96}, and we therefore assume that dimers, once they are formed,
are stable ($i^* = 1$).  Moreover, in the lack of reliable information we
assume that dimers are immobile.  We note that the reported value for the
self-diffusion energy barrier is rather low (0.04 eV)~\cite{stu96,stu94} and
comparable to the energy of optical phonons of Al\,(111) ($0.03 - 0.04$
eV~\cite{egui}). Thus, simulations at room temperature may not be reliable
because the concept of single jumps between nearest neighbor sites is no more
valid. A single optical phonon can furnish enough energy to an adatom for
leaving its adsorption site and diffusing on the flat surface.  At room
temperature the level population of optical phonon is high and the adatoms
have practically no saddle point and migrate freely on the flat surface. We
therefore limited our study to substrate temperatures $T \leq 250$ K.

We adopt periodic boundary conditions, and our rectangular simulation area is
compatible with the geometry of an fcc\,(111) surface. The dimensions of the
simulation area are 1718 $\times$ 2976~\AA$^2$. These dimensions are a
critical parameter and it is important to ensure that the simulation area is
large enough so that artificial correlations of neighboring cells do not
affect the growth patterns.  The mean free path $\lambda$ of a diffusing
adatom before it meets another adatom with possible formation of a nucleation
center or is captured by existing islands should be much smaller than the
linear dimension of the simulation cell~\cite{sch95}. Since $\lambda$ is
proportional to $(D/F)^{1/6}$~\cite{Scaling1}, we find that (with $F = 0.08$
ML/s) $\lambda \sim 50$ \AA~ for $T$ = 50 and gets as large as $\sim 10^3$
\AA~ for $T$ = 250 K. We see that our cell is large enough (for the imposed
deposition rate) for $T \leq 150 K$, whereas at higher temperatures the
dimensions of the cell are too small, i.e., for $T > 150 K$ the island density
is influenced by the sizes of the simulation area. Nevertheless, the island
shape is determined by local processes (edge diffusions) and is still
meaningful.

\begin{figure}[tb]
  \leavevmode
  \includegraphics{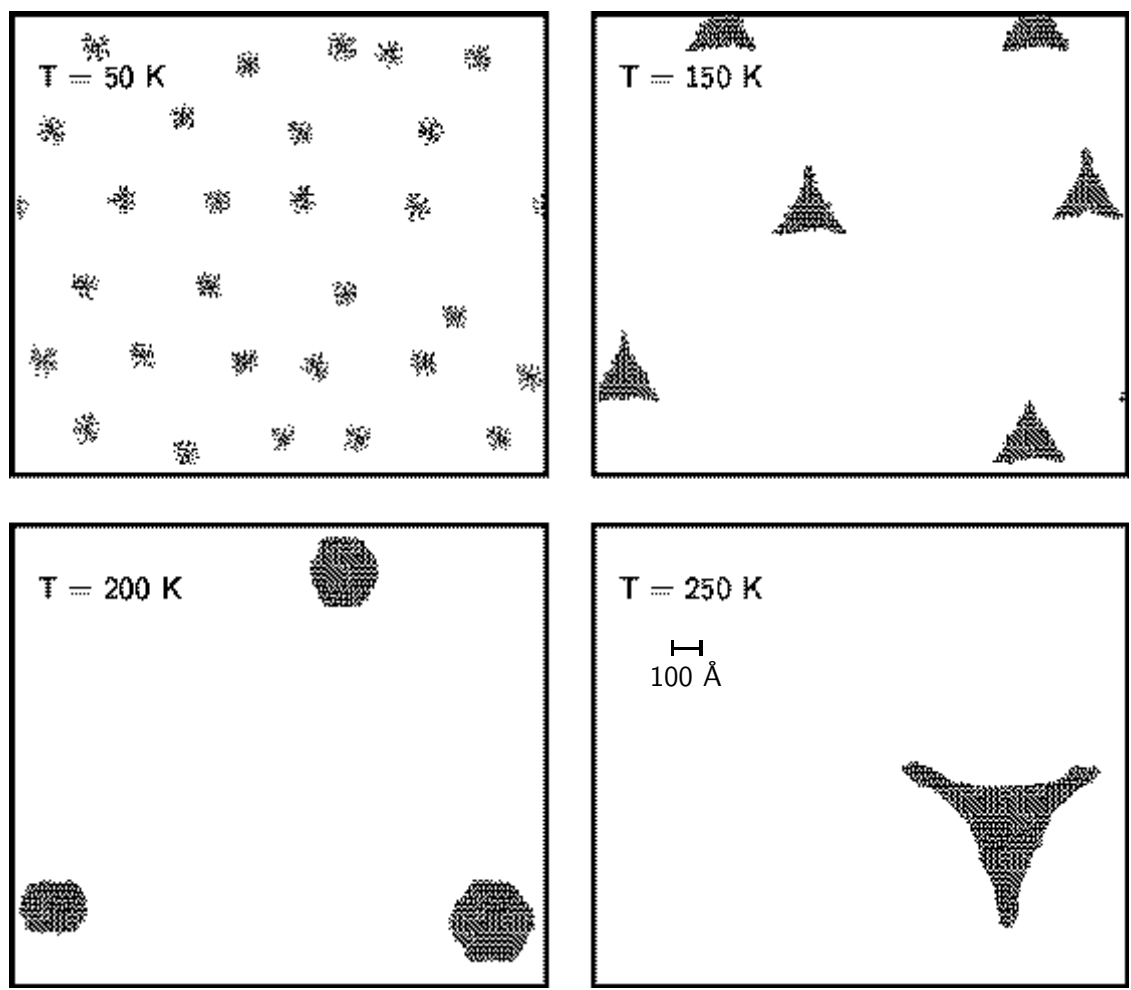}
\unitlength1cm
\begin{picture}(0,0)
\thicklines
\put(9,-9){\line(1,0){0.38}}
\put(9,-9.1){\line(0,1){0.2}}
\put(9.38,-9.1){\line(0,1){0.2}}
\put(8.7,-9.5){\small{\textsf{100 \AA}}}
\end{picture}
\vspace*{13.0cm}
\caption{
A surface area of (1718 $\times$ 1488)~\AA$^2$~ (half of the
  simulation area) at four different substrate temperatures.
The deposition rate was 0.08 ML/s and the coverage in each picture 
is $\Theta$ = 0.04 ML.}
\label{fig.2}
\end{figure} 
In the KMC program two additional insights extracted from the DFT
calculations are included: $(i)$ the attractive interaction between steps and
single adatoms, and $(ii)$ the fact that diffusion processes take place via
different mechanisms (hopping or exchange). Particularly the second point
plays an important role in our investigation. In several KMC simulations of
epitaxial growth the attempt frequency of the diffusion rate has the same
value for all the processes, and this value lies usually in the range of a
typical optical phonon vibration or the Debye frequency. However, this
assumption may not be right. First, processes with larger activation barriers
may have a larger attempt frequency than processes with smaller energy
barriers. This is a consequence of the compensation effect briefly described
in Section~\ref{sec:atomistic}. Moreover, processes as hopping and exchange
that involve a different number of particles and different bonding
configurations may also be characterized by different attempt frequencies.
This has been observed ~\cite{bas78,ayr74,wan89,kel91} for several systems
(Rh, Ir, Pt) and implies that the attempt frequency for exchange diffusion can
be larger by up to two orders of magnitude than that for hopping.
\begin{figure}[b]
\unitlength1cm
\begin{center}
   \begin{picture}(7,6)
      \includegraphics{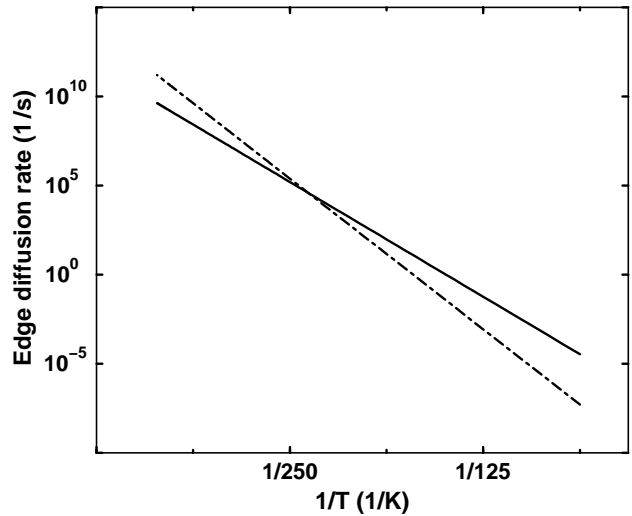}
   \end{picture}
\end{center}
\caption{
Temperature dependence of the edge diffusion rates for atom
  diffusion along the \{100\}-faceted step by hopping with $\Gamma_0$ = $2.5 \times
  10^{12}$ s$^{-1}$
  (solid line), and along the \{111\}-faceted step by exchange with $\Gamma_0$ = $2.5 \times 10^{14}$ s$^{-1}$
  (dash-dotted line).}
\label{fig4}
\end{figure}
For Al surfaces calculations with the embedded atom method~\cite{liu91} showed
a difference of prefactors of one order of magnitude. 

The results of the {\it ab initio} KMC simulations shown in Fig.~\ref{fig.2}
are for a coverage of $\Theta = 0.04$ ML. When the growth temperature is 50~K
the shape of the islands is highly irregular and indeed fractal. Adatoms which
reach a step cannot leave it anymore and they cannot even diffuse along the
step.  Thus, at this temperature ramification takes place into random
directions, and island formation can be understood in terms of the hit and
stick model~\cite{san83}.  At $T$ = 150~K the island shapes are triangular
with their sides being \{100\}-faceted steps. Increasing the temperature to
$T$ = 200~K a transition from triangular to hexagonal shape occurs and for $T$
= 250~K the islands become triangular again.  However, at this temperature
they are mainly bounded by \{111\}-faceted steps.

\begin{figure}[tb]
  \leavevmode 
\includegraphics{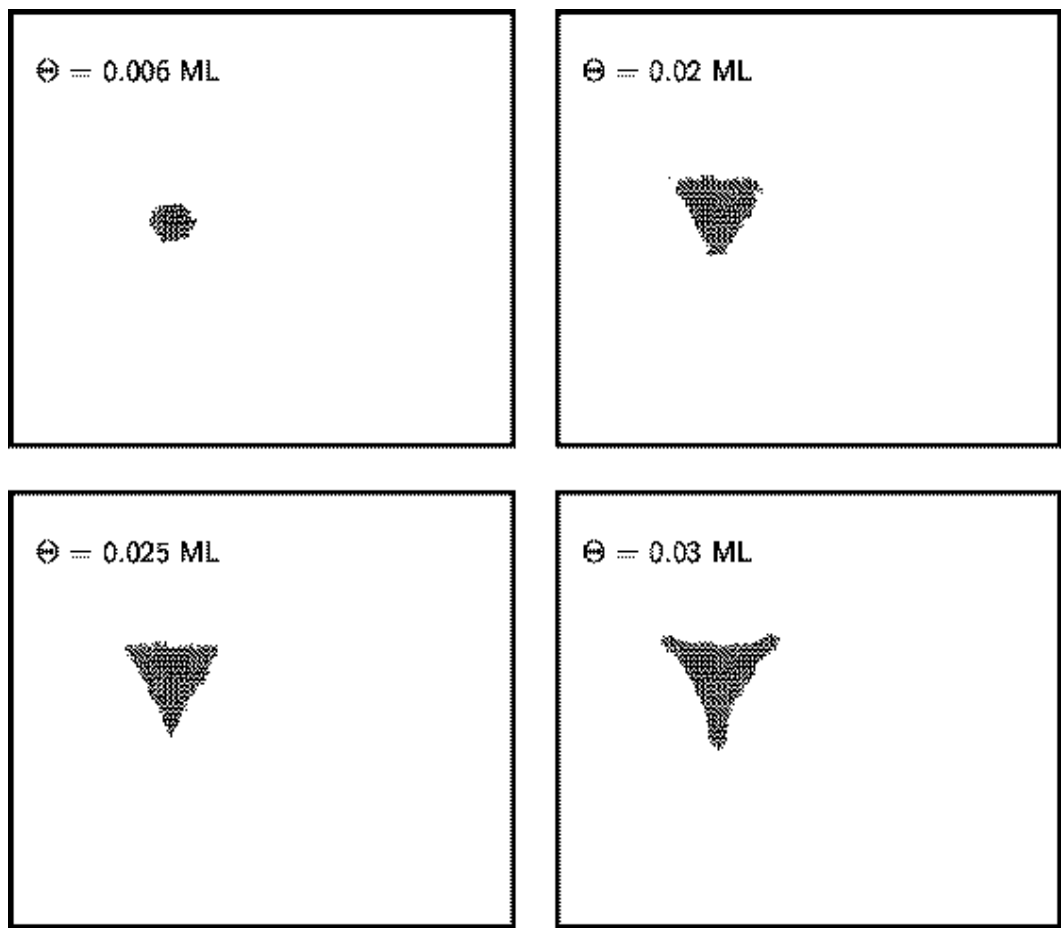} 
\vspace*{13cm}
\caption{
Shape of the islands at $T$ = 250 K as they develop with time (or
coverage). The snapshots refer to
  $\Theta = 0.006$ ML, $\Theta = 0.02$ ML, $\Theta = 0.025$ ML, and $\Theta = 0.03$ ML.
The section of the simulation cell that is shown is 1718 $\times$ 1488~\AA$^2$
and the deposition rate is 0.08 ML/s.}
\label{evolu}
\end{figure} 
To understand the island shapes in the temperature regime between
150 and 250~K we consider the mobility of the adatoms along the steps (at such
temperatures the adatoms at the step edges cannot leave the steps): The lower
the migration probability along a given step edge, the higher is the step
roughness and the faster is the speed of advancement of this step edge. As a
consequence, this step edge shortens and eventually it may even disappears.
Since diffusion along the densely packed steps on the (111) surface (the
\{100\} and \{111\} facets) is faster than along steps with any other
orientation this criterion explains the presence of islands which are mainly
bounded by \{100\}- or \{111\}-faceted steps. The same argument can be
extended to the diffusion along the two close-packed steps and applied to the
triangular islands at $T$ = 150~K, where the energy barrier for the diffusion
along the \{111\} facet is larger and thus the \{100\}-faceted steps survive
so that triangular islands with \{100\} sides are obtained. By considering the
energy barriers we would expect only these islands, until the temperature
regime for the thermal equilibrium is reached.  However, as noted in
Section~\ref{sec:atomistic}, the diffusion rates of adatoms are not only
governed by the energy barrier but also by the effective attempt frequency.
For Al/Al\,(111) the effective attempt frequencies have not been calculated,
but the analysis of Ref.~\cite{stu96} proposes that the exchange process
should have a larger attempt frequency than the hopping process.  The results
displayed in Fig. ~\ref{fig.2} are obtained with $1.0 \times 10^{12}$ s$^{-1}$
for the diffusion on the flat surface, $2.5 \times 10^{12}$ s$^{-1}$ for the
jump along the \{100\}-faceted step, and $2.5 \times 10^{14}$ s$^{-1}$ for the
exchange along the \{111\}-faceted step.  These effective attempt frequencies
are the only input of the KMC not calculated explicitly by DFT, but were
estimated from the theoretical PES as well as from experimental data for other
systems.  In Fig.~\ref{fig4} the edge diffusion rates along the two steps are
plotted as a function of the reciprocal temperature.  At lower temperatures
the energy barrier dominates the diffusion rate but at $T$ = 250 K the attempt
frequencies start to play a role and lead to faster diffusion along the
\{111\} facet than along the \{100\} one.  Thus, the latter steps disappear
and only triangles with \{111\}-faceted sides are present. The roughly
hexagonally shaped islands at $T$ = 200 K are a consequence of the equal
advancement speed for the two steps at that temperature.  Obviously, the
temperature dependence of the growth shapes found in Fig.~\ref{fig.2} is
crucially determined by the ratio of the two diffusivities and in particular
by the temperature at which the two lines of Fig.~\ref{fig4} cross.  If the
difference were only one order of magnitude, the crossing would be at a
temperature that is too high ($500$ K).  The formation of fractals
(Fig.~\ref{fig.2}, upper left) and of \{100\}-faceted step triangles would
still occur.  Obviously, the importance of the attempt frequencies should
receive a better assessment through accurate calculations, and work in this
direction is in progress.

A peculiarity of the triangular islands in Fig.~\ref{fig.2} is that they
exhibit concave sides. In order to understand this behavior we examine the
evolution of the island shape for the deposition at $T$ = 250 K. The results
are collected in Fig.~\ref{evolu}.  At very low coverage the islands are
roughly hexagonal and upon successive deposition they evolve into a nearly
triangular shape. The longer sides are formed by straight \{111\}-faceted step
edges but short \{100\}-faceted edges can still be identified, at least for
$\Theta \leq 0.01$ ML.  The latter edges become rougher and progressively
disappear. For $\Theta = 0.025$ ML the sides are still nearly straight, but at
$\Theta = 0.03$ ML the concavities appear. The corners of the triangles seem
to increase their rate of advancement during deposition. The effect can be
understood on the basis of competition between adatom supply from the flat
surface and mass transport along the sides. The adatom concentration field
around an island exhibits the steepest gradient close to the corners, and the
corners of the islands receive an increased flux of adatoms. When the sides of
the islands are not too long, this additional supply of adatoms is compensated
by the mass transport along the steps, i.e., the adatoms have a high
probability to leave the region around the corners before the arrival of the
successive adatom. For $\Theta = 0.025$ ML this scenario still seems to be
true, while at $\Theta = 0.03$ ML the island edges are longer and the mass
transport along the sides is not able to compensate the additional supply of
particles at the corners. That means that the probability for a particle to
leave the corner region and to move along the island edge before being reached
by another particle decreases considerably, and the corners start to grow
faster than the sides of the triangles so that the concave shape develops.

\subsection{Ag\,(111)}}
\label{sec:ag111}
\subsubsection{The influence of strain on surface diffusion}
\label{sec:ag111str}
Growth of one material on a different material is of particular interest for a
number of technological applications. In such a heteroepitaxial system with
usually different lattice constants the material to be deposited is under the
influence of epitaxial strain.  Growth of Ag on Pt\,(111) and Ag on a thin Ag
film on Pt\,(111) has been the focus of a number of recent studies
\cite{ker95,roe_nature93,bru_nature94}, and with a lattice mismatch of
$4.2\,\%$ it serves as an ideal system that can provide important information
about the effects of strain during growth.  We will particularly discuss how
strain affects the surface diffusion barrier.

Only few theoretical studies of the effect of lattice mismatch on the
diffusion barrier are present in the literature.  For a metallic system we are
only aware of results for Ag on Ag\,(111) where the authors of Ref.
\cite{ker95} find in an EMT calculation that the diffusion barrier increases
under tensile strain and decreases under compressive strain.

Here, we present first principle calculations (more details are given in Ref.
\cite{rat97}) where we study systematically the dependence of the diffusion
barrier on the lattice constant for Ag on Ag\,(111)~\cite{ker95}.  In the
range of $\pm 5\,\%$ strain the DFT results exhibit a linear dependence with a
slope of $0.7$ eV as it is illustrated in Fig.~\ref{Strain_barrier}.
\begin{figure}[t]
\unitlength1cm
\begin{center}
   \begin{picture}(7,6)
      \includegraphics{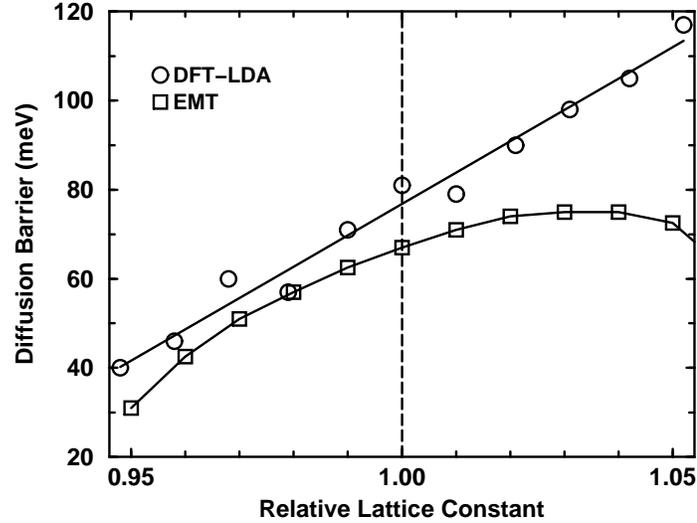}
   \end{picture}
\end{center}
\caption{
Diffusion barrier (in meV) for Ag on Ag\,(111) as a function of strain.
The circles are DFT-LDA results from Ref. \protect{\cite{rat97}}
and the squares are EMT results from Ref. \protect{\cite{ker95}}.}
\label{Strain_barrier}
\end{figure} 
The calculated diffusion barrier for the unstrained system,
$E_{\rm d}^{\rm{\mbox{\scriptsize{Ag-Ag}}}} = 81$~meV, is in good agreement
(within the error margins of the experiment and the calculations) with the
scanning tunneling microscopy (STM) results of $E_{\rm
  d}^{\rm{\mbox{\scriptsize{Ag-Ag}}}} = 97$~meV. The accordance between
experiment and theory extends to the system Ag/Pt\,(111) and Ag/1ML
Ag/Pt\,(111). These results are summarized in \mbox{Table
  \ref{Ag_Pt_barriers}.} In Fig.~\ref{Strain_barrier} the DFT-LDA results are
compared to those of an EMT study~\cite{ker95}.  The EMT results exhibit a
linear dependence only for very small values of strain ($\pm2\,\%$) and the
diffusion barrier starts to decrease for values of misfit larger than $3\,\%$.
\begin{table}[b]
\vspace{.5cm}
\caption{
Diffusion barriers (in meV) for Ag on Pt\,(111), Ag on
one monolayer (ML) Ag on Pt\,(111), and Ag on Ag\,(111).}
\begin{tabular}{lcccccc}
\hline
System & \hspace{.0cm} & Experiment \protect{\cite{ker95}} & \hspace{.0cm} &
EMT \protect{\cite{ker95}} & \hspace{.0cm} & DFT \protect{\cite{rat97}} \\
\hline
Ag/Pt\,(111) & & 157 &  & 81 & & 150 \\
Ag/1ML Ag/Pt\,(111) & & 60 & & 50 & & 65 \\
Ag/Ag\,(111) &  & 97 & & 67 & & 81 \\
\hline
\end{tabular}
\label{Ag_Pt_barriers}
\end{table} 
Indeed, it is plausible that a decrease of the diffusion barrier
occurs when the atoms are separated far enough that eventually bonds are
broken.  However, as our DFT-LDA results show, for Ag/Ag\,(111) this happens
at values for the misfit that are larger than $5\,\%$.  Additionally, when
comparison with experiment is possible [i.e., Ag on Ag\,(111), and Ag on a
monolayer Ag on Pt\,(111)] the EMT results are off by a factor that varies
from 1.2 to 2.

The DFT results in Fig.~\ref{Strain_barrier} were obtained with the LDA for
the exchange-correlation functional and test calculations show that GGA
increases the diffusion barrier by no more than $5 - 10 \,\%$.  The
\,(111)-surface is a closed packed surface with a very small surface
corrugation and since LDA and GGA results on this surface do not show
significant differences for the diffusion barrier (as long as the mechanism is
hopping and not exchange) it is plausible to assume that both the LDA and the
GGA are good approximations for the exact exchange-correlation functional.
This is also true for Pt/Pt\,(111)~\cite{boi96} and Ag on
Ag\,(100)~\cite{yu96}.  The general trend of an increasing energy barrier for
hopping diffusion with increasing lattice constant is quite plausible (for
exchange diffusion see Section~\ref{sec:ag100str}).  Smaller lattice constants
correspond to a reduced corrugation of the surface, and as result the atom is
not bonded much stronger at the adsorption sites than at the bridge site.  In
contrast, when the surface is stretched the corrugation increases and the
adsorption energy at the three-fold coordinated hollow sites increases. This
picture will change when the strain is so large that bonds are broken and then
it is expected that the hopping diffusion barrier will start to decrease again
at very large tensile strain.

It is worth noting that the diffusion barrier for Ag on top of a pseudomorphic
layer of Ag on Pt\,(111) is substantially lower than it is for Ag on
Ag\,(111). A question that arises is whether this reduced diffusion barrier is
a result of the compressive strain or should be ascribed to electronic
rearrangements induced by the Pt substrate.  The diffusion barrier for Ag on
Ag\,(111) with a lattice constant that is compressed to the value of the
lattice constant for Pt is $E_{\rm d}^{\rm{\mbox{\scriptsize{Ag-Ag}}}} =
60$~meV while that for Ag on Pt\,(111) (also with the Pt lattice constant of
3.92 \AA \, obtained from DFT) is $E_{\rm
  d}^{\rm{\mbox{\scriptsize{Ag-Ag/Pt}}}} = 65$~meV. The agreement of these two
values suggests that the reduction of the diffusion barrier for Ag on a layer
of Ag on Pt\,(111) is mainly a strain effect and that the diffusion barrier on
top of a layer of Ag is essentially independent of the substrate underneath.

Brune {\it et al.} \cite{ker95} also measured the island densities of Ag on
two monolayers (ML) of Ag on Pt\,(111) and found that the island density is
much larger than it is for Ag on just one ML of Ag on Pt\,(111). But the
reason for this increased island density is not a larger barrier for surface
diffusion. The second layer of Ag on Pt\,(111) reconstructs in a trigonal
network where domains with atoms in the fcc and hcp site alternate
\cite{bru94}. This reconstruction occurs either during growth with high enough
adatom mobility or upon annealing and it can be concluded that this trigonal
network is the equilibrium structure.  The periodicity of these domains is
approximately two domain boundaries for every 24 atoms. This can be understood
very well with purely geometrical arguments because the lattice mismatch is
$4.2\,\%$ and every domain boundary implies that there is half of an Ag atom
less so that the domain network provides an efficient mechanism to relief
epitaxial strain.  The barrier to diffuse across such a domain wall appears to
be rather high and domain walls act as repulsive walls so that the island
density is determined by the defect density and not the barrier for self
diffusion.  It is not clear however why this domain network is formed only
after 2 ML Ag have been deposited and not already upon completion of the first
Ag layer.

To answer this question the adsorption energy of an adatom in the fcc and in
the hcp site were compared by Ratsch {\it et al.}~\cite{rat97}.  Calculations
were carried out with a ($1 \times 1$) and a ($2 \times 2$) cell and slab
thicknesses of up to 5 layers. It was found that the fcc site is energetically
more favorable than the hcp site in all cases. The energy differences between
the two adsorption sites for the first and second Ag layer on top of Pt\,(111)
and Ag on Ag\,(111) are summarized in Table~\ref{ad_sites_Ag}.
\begin{table}[tb]
\vspace{0.5cm}
\caption{
Energy difference $\Delta E = -(E_{\rm fcc} - E_{\rm hcp})$ 
between the 
total energies of the two adsorption sites for Ag on Pt\,(111) and 
Ag on Ag\,(111).}
\begin{tabular}{lccccc}
\hline
System &\hspace{1.5cm} & $\Delta E$ (in meV) & \hspace{1.5cm} & $\Delta E$ (in meV) \\
& & $1 \times 1$ cell & & $2 \times 2$ cell \\
\hline
Ag/Pt(111) & & 30 &  &50 \\
Ag/1ML Ag/Pt(111) & & $\sim$ 0 & & $<$5 \\
Ag/Ag(111) & & $<$10 & & $<$10 \\
\hline
\end{tabular}
\label{ad_sites_Ag}
\end{table} 
During deposition of the first layer the fcc site is energetically
preferred compared to the hcp site, and the film growth pseudomorphically with
the atoms in the fcc sites.  However, after the first layer has been
completed, the total energies of these two sites are almost indistinguishable
so that the system is not prevented from reconstructing in the described
domain network to relieve epitaxial strain.

\subsubsection{The role of antimony as a surfactant}
\label{sec:surf}
As already mentioned, Ag atoms deposited on Ag(111) form ``mountains'' (or
``mounds'') as it has been seen by RHEED~\cite{suz88}, x-ray reflectivity
experiments~\cite{veg92}, and STM~\cite{Meyer,bro95,vri94} for a wide range of
temperatures.  The situation changes completely when small amounts of antimony
(0.2 ML) are deposited on the surface and the growth mode becomes two
dimensional~\cite{Tersoff94,vri94}.

The natural question that arises is: What is the action of Sb? These
contaminants are called {\it surfactants}, although this term may be
misleading. In the original definition a surfactant should reduce the surface
energy. Particularly for metallic systems, however, the contaminants rather
affect the kinetics of the processes and change the growth mode.

\begin{table}[b]
\caption{
Adsorption energies (in eV/atom) of Sb on Ag\,(111).}
\label{ads.sb.111}
\begin{tabular}{lcccccc}
\hline
& \hspace{1.8cm} & $2 \times 2$ & \hspace{1.8cm} & 
($\sqrt{3} \times \sqrt{3}$)R30$^o$ & \hspace{1.8cm} & $1 \times 1$\\
\hline
$E^{\rm sub}_{\rm ad}$ & & 4.37 & & 4.49 & & - \\
$E^{\rm fcc}_{\rm ad}$ & & 3.34 & & 3.26 & & 3.22 \\
$E^{\rm sublayer}_{\rm ad}$ & & 3.45 & & 3.41 &  & 2.71 \\
\hline
\end{tabular}
\end{table} 
In Section~\ref{sec:critical} we have described some possible
scenarios for the action of these surfactants.  Clearly, different surfactants
work by different mechanisms (or a combination of them).  However, as already
pointed out in Section~\ref{sec:critical} a good surfactant has to satisfy one
essential requirement: It has to stay on the surface.  To investigate this
crucial point, Oppo {\it et al.}~\cite{scheff96,opp93} carried out DFT-LDA
calculations for the adsorption energies of Sb atoms located in different
positions on Ag\,(111).  These calculations have given important insight into
the action of Sb during growth.  The evaluated binding energies for Sb are
collected in Table~\ref{ads.sb.111} for different coverages.  It is clear that
the substitutional site is greatly favored with respect to on-surface fcc
adsorption and sublayer adsorption for all coverages considered ($\Theta_{\rm
  Sb}$ = 1/4, 1/3, and 1 were considered)).  Sb is thus expected to be
adsorbed in substitutional sites at not too high Sb coverages and this
geometry is shown in Fig.~\ref{sb.ag.111}.
\begin{figure}[t]
\unitlength1cm
\begin{center}
   \begin{picture}(10,3.5)
      \includegraphics{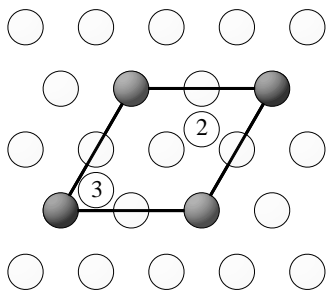}
   \end{picture}
\end{center}
\caption{
Adsorption sites of Sb on Ag\,(111) considered in the calculations of
Ref.~\protect\cite{opp93}.}
\label{sb.ag.111}
\end{figure} 
An incorporation of the contaminant into the Ag bulk can be ruled
out because its energy cost is too high. The Sb atoms segregate to the surface
and do not incorporate into the bulk, since its sizes are somewhat too large
for a bulk vacancy, but appropriate to fit into a surface vacancy.

The DFT results clarify the energetical ranking for the adsorbed Sb atom, but
the understanding of its action on the Ag atoms requires the determination of
its influence on the Ag adsorption.  Two possible locations of the Ag adatom
on the substitutional Sb-covered surface are considered: a {\it near} site and
a {\it far} site, depending on whether Ag and Sb are nearest neighbors or not.
These two sites are labeled 3 and 2 in Fig.~\ref{sb.ag.111}, and their
calculated adsorption energies are 1.99 eV/atom and 2.02 eV/atom,
respectively.  The Ag adatom prefers energetically to sit on the clean portion
of the surface ({\it clean} site with $E^{\rm Ag}_{\rm ads} = 2.41$ eV/atom),
whereas close to an incorporated Sb atom it favors the {\it far} site. Thus,
Ag adatoms avoid the vicinity of the Sb atoms.

A schematic summary of the action of Sb during growth of Ag on Ag\,(111)
is given in Fig.~\ref{surf.act}.
\begin{figure}[b]
\unitlength1cm
\begin{center}
   \begin{picture}(10,9.0)
      \includegraphics{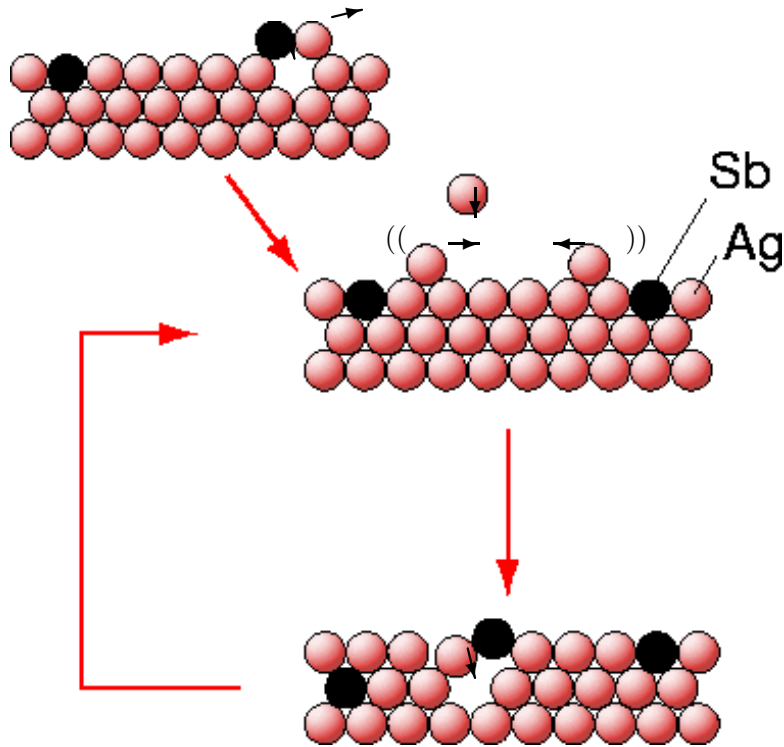}
      \thicklines
      \put(2.25,8.4){\vector(1,-4){.1}}
      \put(2.85,8.56){\vector(4,1){.4}}
      \put(4.75,6.3){\vector(0,-1){.4}}
      \put(4.4,5.55){\vector(1,0){.4}}
      \put(6.2,5.55){\vector(-1,0){.4}}
      \put(3.55,5.5){\small{((}}
      \put(6.75,5.5){\small{))}}
%
      \put(4.65,0.16){\vector(1,-4){.1}}
   \end{picture}
\end{center}
\caption{
Sketch of the action of Sb on Ag\,(111).}
\label{surf.act}
\end{figure} 
The substitutional antimony atoms act as repulsive centers for
the silver adatoms (upper panel), whose mean free path is reduced by the
presence of the contaminants (middle panel). Thus, small-sized Ag islands
appear on the surface with an increased density.  Moreover, the antimony atoms
are expected to affect the form of the Ag islands, and the islands will grow
in directions where they can avoid to get close to Sb centers. Finally, just
before the Ag-Sb surface alloy layer is covered by the newly deposited Ag
layer, the Sb atoms find themselves in the disfavored subsurface configuration
(see, for instance, Table~\ref{ads.sb.111}), and segregate to the new surface
layer (Fig.~\ref{surf.act}, lower panel). The alloy layer is now reestablished
and the process can start again. The combination of all these features yields
layer-by-layer growth.  It is interesting to note that according to
Eq.~(\ref{eq:lbl}) only an increase of island density by a factor of 2 is
necessary to induce two dimensional growth. Vrijmoeth {\it et
  al.}~\cite{vri94} report a much higher increase of the island density (a
factor of $\sim$\,6) of the annealed surface in the presence of Sb.

A recent theoretical study~\cite{liu95} where a rate-equation analysis has
been combined with Monte Carlo simulations has confirmed the main idea
proposed in the work of Oppo {\it et al.}~\cite{scheff96,opp93}. The main role
of the Sb repulsive network is to reduce the mean free path of the Ag adatoms,
and as a result the island density is enhanced. The changed character of
diffusion leads to new scaling relations in which the island density shows a
strong dependence on the impurity concentration.

\subsection{Microscopic processes at Al\,(100)}         
\label{sec:al100}

Self-diffusion at flat regions of close-packed surfaces is typically
considered to proceed by series of hops between adjacent equilibrium
adsorption sites.  For a fcc\,(100) surface an adatom in a fourfold hollow
site moves over the twofold bridge site to a neighboring fourfold hollow site
[see Fig.~\ref{more_exchange}(a)].  After some time an adatom has visited most
fourfold hollow surface sites, and the visited sites form a ($1 \times 1$)
pattern.  An alternative mechanism for surface diffusion is atomic exchange
where the adatom moves by displacing a neighboring surface atom. On a
fcc\,(100) surface the exchange process occurs along the \,[010\,] and
[\,001\,] directions.  The transition state [see Fig.~\ref{more_exchange}(b)]
may be described as a dimer, consisting of the adatom and the lift-up surface
atom located above a surface vacancy.  Subsequently the displaced surface atom
becomes a new adatom in a next-nearest-neighbor fourfold site.  When this
diffusion mechanism is active, an adatom visits only every second fourfold
hollow site at the surface, and the visited sites form a c($2 \times 2$)
pattern.  This difference in the patterns of visited sites represents clear
evidence of the active mechanism. Indeed, in this way self-diffusion by
exchange has been experimentally observed by Kellogg and Feibelman for
Pt\,(100)~\cite{kel90} and by Chen and Tsong for Ir\,(100)~\cite{che90}.

A theoretical study of exchange diffusion was performed by
Feibelman~\cite{exchange} for Al\,(100).  Feibelman performed DFT calculations
and predicted a rather high energy barrier for hopping ($E_{\rm d}^{\rm hop} =
0.65$~eV) in conflict with experimental estimates.  Inspired by ideas of
Pandey~\cite{pand86} Feibelman realized that at surfaces it is more
appropriate to think of self-diffusion in terms of making and breaking of
chemical bonds rather than in terms of a hard sphere rolling over a bumpy
plane. In fact, his calculations of the exchange mechanism show that its
barrier is substantially smaller (by $\approx 0.45$~eV) than that of the
hopping process.  Thus, he predicted that for Al\,(100) self-diffusion
proceeds by atomic exchange. The process is caused by the noticeable covalency
of aluminum, which can form directional bonds (by $sp$ hybridization) at
certain atomic geometries.  In the transition state of exchange diffusion at
Al\,(100) [see Fig.~\ref{more_exchange}(b)] each atom of the dimer that sits
over the vacant site forms three chemical bonds, two with surface Al atoms and
one with the other atom of the dimer.  Inspection of the electron
density~\cite{exchange} revealed the formation of directional bonds and that
the coordination of the surface atoms at the transition state can be described
as three-fold.  In other words: At the transition state of hopping diffusion
the adatom is two-fold coordinated, while at the transition state of the
exchange diffusion it is three-fold coordinated.  For a group III atom it thus
appears plausible that the latter transition state has a lower energy.  We
note in passing that recent embedded atom calculations have questioned the
importance of the exchange mechanism at Al\,(100)~\cite{liu91}, but DFT
calculations of Stumpf and Scheffler~\cite{stu96} fully confirmed the scenario
proposed by Feibelman.

These studies are predictions with only indirect experimental support, namely
the observation that the diffusion barrier is low.  Only for two other
fcc\,(100) surfaces exchange diffusion has been identified [Pt\,(100) and
Ir\,(100)].  Apparently it does not occur at other fcc\,(100) transition-metal
surfaces. Recent calculations have shown that the mechanism that stabilized
the exchange diffusion at the late $5d$ transition metals is significantly
different from the covalency mechanism at Al\,(100).  In
Section~\ref{sec:ag100str} we will summarize these results.

\subsection{Ag\,(100)}
\label{sec:ag100}
\subsubsection{Microscopic processes}
\label{sec:ag100mic}

For silver it is known that growth on the (111) surface is three dimensional
because a noticeable step-edge barrier exists that has been estimated by
various authors using STM~\cite{Meyer,bro95} to be between 0.1 and 0.15 eV.
This additional step-edge barrier explains why the growth of silver
perpendicular to the (111) surface proceeds in a three-dimensional mode.

In order to analyze why the growth of the Ag\,(100) surface is qualitatively
different (two dimensional) Yu and Scheffler~\cite{yu96} performed DFT
calculations within the LDA and the GGA, and the results are listed in
Table~\ref{act.en.100}.  In contrast to the above discussion for Al\,(100)
self-diffusion on Ag\,(100) proceeds by hopping. Exchange diffusion has an
energy barrier which is about 0.3 eV higher.
\begin{table}[b]
\caption{
Energy barriers $E_{\rm d}$ (in eV) for different processes on Ag\,(100)
  calculated within the LDA (GGA).}
\begin{tabular}{lcccc}
\hline
process & \hspace{2.8cm} & mechanism & \hspace{2.8cm} & $E_{\rm d}$ (eV) \\
\hline
 flat Ag(100) & & hopping & & 0.52 (0.45) \\
 flat Ag(100) & & exchange & & 0.93 (0.73) \\
 step $\|$ & & hopping & & 0.30 (0.27) \\
 step $\perp$ descent. & & exchange & & 0.52 (0.45) \\
 step $\perp$ descent. & & hopping & & 0.70 (0.55) \\
\hline
\end{tabular}
\label{act.en.100}
\end{table}

However, according to the DFT calculations diffusion across a descending
close-packed step takes place via the exchange mechanism (cf.
Table~\ref{act.en.100}). The calculated activation energy for the step-down
motion by exchange is (within the numerical accuracy) identical to that of the
hopping diffusion at the terrace.  Figure~\ref{tot.en.dif.10} shows the total
energy curves along the diffusion paths for hopping and exchange diffusion.
\begin{figure}[b]
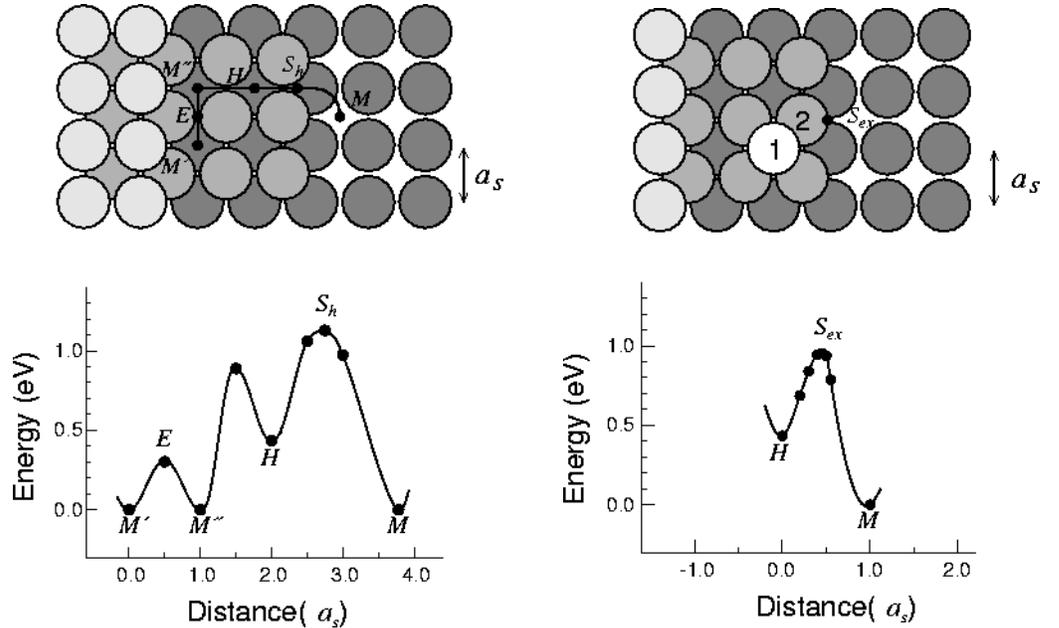

  \leavevmode 
  \includegraphics{Fig21}
  \includegraphics{Fig22}
\vspace*{7.5cm}
\caption{
Diffusion path (upper panel) and total energy (lower panel) of an
Ag adatom diffusing across a descending step by hopping (left side) and
exchange process (right side). The values of the energy have been 
calculated within the LDA. For the
exchange mechanism the total energy is plotted as function of the distance of
the step edge atom 2 from the undistorted step edge.}
\label{tot.en.dif.10}
\end{figure}

The low energy barrier of the exchange diffusion can be understood by
considering the geometries along the reaction pathway: Atom 2 moves out of the
step edge and the process is energetically stabilized because adatom 1 keeps
in contact with atom 2 and finally occupies the original site of atom 2
(Fig.~\ref{tot.en.dif.10}, right side).  The transition state for the exchange
process is near the bridge site formed by two step-bottom atoms on the lower
terrace ($S_{ex}$ in Fig.~\ref{tot.en.dif.10}, right side).  Inspection of the
geometry at $S_{ex}$ (see Ref.~\cite{yu96} for details) shows that the local
coordination of atoms 1 and 2 remains high.  At $S_{ex}$ each of theses atoms
is five-fold coordinated.  This finding that there is no additional energy
barrier provides a natural explanation for the smooth 2-D growth of Ag\,(100).
By analyzing the STM images of Ag(100) during deposition Zhang {\em et
  al.}~\cite{Zhang96} gave an estimate of $0.025\pm0.005$~eV for the
additional step-edge barrier in good agreement with the DFT results.  We also
note that the coordination number arguments employed above are valid in
general, thus we expect that our finding that step-down diffusion at Ag\,(100)
proceeds by the exchange mechanism also applies to other noble and other
fcc\,(100) transition-metal surfaces.

Yu and Scheffler also studied the diffusion parallel to step edges and the
step formation energies~\cite{yu97}. They find that diffusion parallel to step
edges is faster than that at flat surface regions and that the closest-packed
step, the $\langle 110 \rangle/\{111\}$ step dominates. Thus, it is predicted
that step edges are rather straight and that in equilibrium islands have an
approximately square shape. To be precise, the size is that of an octagon with
long $\langle 110 \rangle/\{111\}$ and shorter $\langle 100 \rangle/\{110\}$
steps.  From the step formation energy the length ratio was determined to be
10:3.

\subsubsection{The influence of strain on surface diffusion}
\label{sec:ag100str}

As noted above, self-diffusion on fcc\,(100) surfaces may proceed by the
hopping process (as for Ag) or by exchange (Pt, Ir, and probably also Al).  Yu
and Scheffler~\cite{yu97} recently predicted that exchange diffusion should
occur on Au\,(100) and strained Ag\,(100), and two mechanisms were emphasized
that stabilize the exchange over the hopping mechanism: $i)$ the tensile
surface stress, and $ii)$ the correlation of bond strength and local
coordination.

For Au\,(100) the rather low energy of the exchange-diffusion transition state
($E_{\rm d}^{\rm ex} = 0.65$~eV, $E_{\rm d}^{\rm hop} = 0.83$~eV) could be
correlated with its geometry (compare Fig.~\ref{more_exchange}(b)): The dimer
is only 1.29~\AA{} above the surface layer, which is 37~\%{} closer to the
center of the top layer than the inter-layer spacing in the bulk.  Thus, the
two atoms of the dimer interact with the atoms of the top layer of the
substrate but get rather close to the second layer.  In other words, the
surface likes to attain closer packing which reflects its noticeable tensile
stress~\cite{Fiorentini}.  Indeed, the surface stress of the late $5d$ metals
is significantly higher than that of their $4d$ and $3d$ isoelectronic
elements.  The difference has been traced back to relativistic
effects~\cite{Fiorentini}, which play a noticeable role for the heavier $5d$
metals: The relativistic effects give rise to a contraction and energy
lowering of $s$-states, and as a consequence, the $d$-band moves closer to a
Fermi energy~\cite{Fiorentini}.  Indeed, a relativistic treatment is most
important in order to attain a good description of structural and elastic
properties of $5d$ metals, while it is not important for the $4d$ metals.
Thus, while the significant tensile surface stress of Au\,(100) pulls the
dimer of the exchange transition state ``into'' the surface, it lowers the
energy of the transition state, and enables exchange diffusion. The surface
stress at Ag\,(100) is too weak to have a significant effect.

With respect to the stress a stretched silver film gets more gold-like, and
therefore the two diffusion mechanisms for a strained silver slab were
analyzed in detail.  When the system is under tensile strain the energy
barrier for the exchange diffusion increases, while the barrier for the
hopping diffusion decreases.  For hopping diffusion the trend can be
understood as follows: Smaller lattice constants correspond to a reduced
corrugation of the surface potential, and thus diffusion energy barriers are
reduced~\cite{rat97,ker95}.  In contrast, for a stretched surface the
corrugation increases and the adsorption energy at the four-fold coordinated
hollow sites increases.  The latter reflects the wish to reduce (at least
locally) the strain induced surface stress. The hopping-diffusion transition
state is less affected by the strain than the adsorption site.  For exchange
diffusion this is just the opposite. Here the transition-state geometry reacts
particularly strongly to the tensile stress, and locally the tensile stress is
reduced by the very close approach of the dimer [Fig.~\ref{more_exchange}(b)].
Thus, it is predicted that for pseudomorphic Ag films (with increased parallel
lattice constant) self-diffusion should get noticeably affected by the
exchange mechanism. The results of this study strongly suggest that tensile
surface stress (to be precize, the {\em excess} surface stress, i.e. the
strain derivative of the surface energy) is the main actuator for the exchange
diffusion on fcc\,(100)
surfaces~\cite{yu97}.
}

\fontsize{12}{12.2}\selectfont{

}

\begin{thebibliography}{9999}
\bibitem{units} All equations are noted in Hartree atomic units, i.e.
the unit of length is 1 Bohr = 0.5292 \AA, the unit of energy
is 1 Hartree = 27.2116 eV, and $\hbar = m = e = 1$.
\end{thebibliography}
\end{document}